\documentclass[11pt,french,english,      ]{article}
\usepackage[T1]{fontenc}
\usepackage[latin1]{inputenc}
\usepackage{geometry}
\usepackage{babel}
\usepackage{graphics}

\makeatletter

\providecommand{\LyX}{L\kern-.1667em\lower.25em\hbox{Y}\kern-.125emX\@}

\makeatother
\begin{document}

\title{Parameterization of two-dimensional turbulence using an anisotropic
maximum entropy production principle\\
}

\author{F. BOUCHET{\small \( ^{1,2} \)} \\
\\
{\small \( ^{1} \)Institut Fourier, UMR 5582, Grenoble France}\\
{\small \( ^{2} \)Dipartimento di Energetica {}``Sergio Stecco'',
Universit\`a degli studii di Firenze, Italy}\\
{\small \nonumber }}

\maketitle
Received:

Author version: \date

Author contact : bouchet@dma.unifi.it 

\begin{abstract}
We consider the modeling of the effect of unresolved scales, for two-dimensional
and geophysical flows. We first show that the effect of small scales
on a coarse-grained field, can be approximated at leading order, by
the effect of the strain tensor on the gradient of the vorticity,
which exactly conserves the energy. We show that this approximation
would lead to unstable numerical code. In order to propose a stable
parameterization, while taking into account of these dynamical properties,
we apply a maximum entropy production principle. The parameterization
acts as a selective diffusion proportional to the mean strain, in
the contraction direction, while conserving the energy. We show on
numerical computation that the obtained \foreignlanguage{french}{anisotropic
relaxation equations} give an important predictability improvement,
with respect to Navier-Stokes, Smagorinsky or hyperviscous parameterizations. 
\end{abstract}
\newpage

\section{Introduction}

Turbulent flows are characterized by the non-linear interaction of
a huge number of variables. For very high Reynolds' numbers, the current
or foreseen computational capabilities only permit to describe part
of corresponding degrees of freedom. For instance, for geophysical
flows, the molecular viscosity acts at typical scales of order of
the millimeter, whereas a current typical resolution for meteorological
computations is about 30 to 50 \( km \). The issue of the parameterization
of the effect of unresolved scale on the actually described ones is
thus of crucial importance. 

The predominance of the Coriolis force and the density stratification
give geophysical flows a quasi two-dimensional nature. Their organization
is actually three dimensional, but may be well approximate by layered
models \cite{Pedlosky}. Moreover, at large scale, an inverse energy-cascade
process (from small to large scale), typical of two-dimensional flows,
is observed for geophysical flows. In the following, we address the
issue of small scale turbulence parameterization in both the frameworks
of the two-dimensional incompressible Euler equation, or of the Quasi-Geostrophic
model on the other hand.

Any numerical resolution of the flow dynamics implicitly assumes that
an averaged value is described. The average of the nonlinear terms
should then be expressed in terms of the known averages of the single
quantities. This is the classical closure problem for flow equations
\cite{Prandtl}. In order to clearly specify the problem, one has
to precise the meaning of the average. When a statistically stationary
situation is identified, a very natural choice is an ensemble average
on the stationary distribution. For instance, this would be the case,
for an inverse cascade regime (a small scale energy injection and
a large scale drag force) where a natural ensemble is given by a stationary
probability distribution for this process. The proximity of such a
stationary situation is implicitly assumed, together with other hypothesis,
in classical parameterizations such as the Eddy Damped Quasi Markovian
Model (EDQNM) \cite{Leslie,Orszag,Lesieur} or the \( K-\varepsilon  \)
model \cite{Orszag}. An other approach is to specify the stochastic
properties of the unresolved scales, as for instance in the Direct
Interaction Approximation (DIA) \cite{Kraichnan}. 

However, geophysical flows are characterized by the existence of large
scales structures (jets, cyclone and anticyclones), which break the
self-similarity at the base of the description of the inverse cascade
process. The typical time scales for forcing and dissipation are much
larger than the inertial ones. In order to study such flows, we will
consider two dimensional inviscid models in a freely decaying-turbulence
regime. In such a case, for large times, the flow self-organize in
coherent structures. The final state may be described by the statistical
mechanics of the vorticity (or the potential vorticity) \cite{Robert_Sommeria_91,Miller}.
For such a freely decaying turbulence, and in a situation close to
the equilibrium or to a quasi-stationary state of the inviscid equations,
it would be very natural to consider an ensemble average, compatible
with the observed large scale structure and vorticity distribution.
Such an ensemble average is for instance implicitly assume in the
kinetic approach of Chavanis \cite{Chavanis_Quasilinear}. On the
contrary, in this work, we will be interested in situations far from
equilibrium or from stationary solutions (vortex merging, filamentation).
In such a case, there is no natural statistical ensemble to consider.
Moreover, we will show that for a given flow evolution, the effect
of unresolved scales will be dominated by structures (filaments) which
are strongly correlated with the resolved scales. Given this situation,
we will consider the evolution of locally averaged quantities, on
the grid scale \( l \) (average by homogenization by opposition of
ensemble average).

In section \ref{sec: moyenis=E9e} we recall the equation of motion
and give the formal properties of an average by homogenization. We
then show that the averaged equations will be determined by the knowledge
of a turbulent current. In section \ref{sec:Flux}, we give an analytic
expression of this turbulent current, at leading order in an asymptotic
expansion in power of the ratio of grid scale on the domain scale
\( l/L \). The turbulent current is then expressed in terms of the
averaged velocity strain tensor. This result has been described in
\cite{Bouchet_these} and independently in \cite{Dubos_CRAS,Dubos_Thèse}.
We discuss the validity of this deterministic approximation. Using
numerical integration, we show that this expression is actually a
very good approximation of the actual turbulent current, in situations
of decaying turbulence, far from equilibrium. This deterministic approximation
exactly conserves the energy of the mean flow.

As the strain tensor is of zero trace, its two eigenvalues are opposite.
They respectively correspond to a positive and negative viscosity.
For this reason, this deterministic approximation is not suited to
model the effect of small scales in a numerical code: it would lead
to instability of the scheme. In section \ref{sec : Relaxation_Anisotrope},
we obtain a stable algorithm by determining the turbulent current
which maximize the production of the mixing entropy, while respecting
both the value of the deterministic current in the stable direction
of the strain tensor, and the conservation of energy. Such a maximum
entropy production principle has already been used by Robert and Sommeria
\cite{Robert_Sommeria_92}, but without taking into account for the
anisotropy of the turbulent current. With this new specification,
the obtained equations respect all the usable properties of the deterministic
current, and have the further property to converge towards equilibrium
of the statistical mechanics. In section \ref{sec: Comparaison_numerique},
we compare the performance of these \emph{anisotropic relaxation equations}
with other models, commonly used to parameterize small scale turbulence,
for two-dimensional or geophysical flows: the eddy viscosity model,
the Smagorinsky model \cite{Smagorinsky}, and an hyperviscosity parameterization.
In section \ref{sec: Comparaison_numerique}, the anisotropic relaxation
equation are shown to give the best results, when compared with high
resolution simulation, both for the Euler and for the Quasi-Geostrophic
equations.

In section \ref{sec_Discussion}, we discuss these results and their
interest for geophysical flow applications.

\section{The averaged equations of motion \foreignlanguage{french}{\label{sec: moyenis=E9e}}}

In this section, we recall the two-dimensional incompressible Euler
equations and the Quasi-Geostrophic model. We specify the average
operator for the resolved dynamics and we give the main properties
of the averaged equations of motion.\\

The incompressible Euler equations describe the evolution of the velocity
\( {\bf u} \) of an incompressible inviscid fluid. For 2D flows,
they are equivalent to the equation for the evolution of the vorticity
\foreignlanguage{french}{(\( \omega =\left( \nabla \wedge {\bf u}\right) .{\bf e}_{z} \))}:

\selectlanguage{french}
\begin{equation}
\label{Euler}
\frac{\partial \omega }{\partial t}+{\bf u}.\nabla \omega =0
\end{equation}
\begin{equation}
\label{Dirichlet_Psi}
\omega =-\Delta \psi 
\end{equation}
\begin{equation}
\label{U_psi}
{\bf u}=-{\bf e}_{z}\wedge \nabla \psi 
\end{equation}
The vorticity \( \omega  \) is advected by the divergence-less velocity
\( \bf {u} \), \( \psi  \) is the stream-function. For sake of simplicity,
we will consider these equation on a torus \( D \) (all the quantities
are periodic in both variables, with zero global average). All our
discussion is easily generalizable for any bounded domain \( D \)
(an impermeability condition \( {\bf u}.{\bf n} \), where \( {\bf n} \)
is normal to the boundary of \( D \) would then specify the boundary
conditions).

The Euler equations conserve the energy:

\begin{equation}
\label{Energie_Euler}
E=\frac{1}{2}\int _{D}d{\bf r}\, {\bf u}^{2}=\frac{1}{2}\int _{D}d{\bf r}\, \omega \psi 
\end{equation}
and all the functionals of the form \( C_{f}(\omega )=\int _{D}d{\bf r}\, f(\omega ) \)
for any function \( f \). The Quasi-Geostrophic (QG) equation, which
is the simplest model of geophysical flows \cite{Pedlosky}, are:

\begin{equation}
\label{QG}
\frac{\partial q}{\partial t}+{\bf u}.\nabla q=0
\end{equation}
\begin{equation}
\label{Dirichlet_Psi_QG}
q=-\Delta \psi +\frac{\psi }{R^{2}}
\end{equation}
We note the formal similarity with the Euler equation. The potential
vorticity \( q \) is then advected. The determination of the stream
function, is then given by the inversion of the equation (\ref{Dirichlet_Psi_QG}),
and the velocity is still given by (\ref{U_psi}). The energy for
the QG dynamics is \( E=1/2\int _{D}d{\bf r}\, \left[ {\bf u}^{2}+\psi ^{2}/R^{2}\right] =1/2\int _{D}d{\bf r}\, q\psi  \).
One clearly see the strong analogies between the QG and the Euler
equations. From now on, we will deal only with the Euler equation.
We will precise when the generalization for the QG equation is not
straightforward.\\

Our aim is to obtain the equation describing the evolution of the
spatial average of the velocity and of the vorticity. We define the
average \( \overline{f} \) of any quantity \( f \) by:\[
\overline{f(\bf {r})}=\int _{D}d{\bf r'}\, G^{l}({\bf r}-{\bf r'})f({\bf r'})\]

This operator is the convolution with \( G. \) We suppose that \( \int d{\bf r}\, G({\bf r})=1 \)
and that \( G \) is invariant under any rotation (\( G({\bf r}) \)
only depends on the modulus of \( {\bf r}. \)) The qualitative properties
assumed for \( G \) are to regularize the field at a fixed scale
\( l \). For instance, in section \ref{sec:Flux}, we will use in
the \( 2\pi  \)-periodic domain the kernel \( G^{l} \) whose Fourier
components are \( G^{l}_{\bf {k}}=\frac{1}{\left( 2\pi \right) ^{2}}\exp \left( -\frac{l^{2}k^{2}}{2}\right) ,\, \bf {k}\in Z^{2} \)
(when \( l \) is much smaller than \( 2\pi  \), \( G^{l}({\bf r}) \)
is close to \( \frac{1}{2\pi l^{2}}\exp \left( -\frac{r^{2}}{2l^{2}}\right)  \)).
The average operator verifies the linearity property: \( \overline{f+\lambda g}=\overline{f}+\lambda \overline{g} \)
and the commutation with the differential operators: \( \partial _{i}\overline{f}=\overline{\partial _{i}f} \).
We note that it does not verify the properties \( \overline{\overline{f}}=\overline{f} \)
and \( \overline{f\overline{g}}=\overline{f}\overline{g} \) as would
be the case for an ensemble average. In section \ref{sec:Flux}, we
will show that the dominant contribution of the turbulent current
comes from the non-verification of these two properties.

By averaging (\ref{Euler}), we obtain the averaged Euler equation:
\begin{equation}
\label{Euler_moy}
\frac{\partial \overline{\omega }}{\partial t}+\overline{{\bf u}}.\nabla \overline{\omega }=-\nabla .\bf {J}_{\omega }\, \, \, \rm {with}\, \, \, \bf {J}_{\omega }=\overline{\omega {\bf u}}-\overline{\omega }\, \overline{{\bf u}}
\end{equation}
 \begin{equation}
\label{Dirichlet_Psi_moy}
\overline{\omega }=-\Delta \overline{\psi }
\end{equation}
\begin{equation}
\label{U_psi_moy}
\overline{{\bf u}}=-{\bf e}_{z}\wedge \nabla \overline{\psi }
\end{equation}
We note that the correlation term is the divergence of a current \( \bf {J_{\omega }} \),
that we will call the turbulent current. We note that the average
Euler equations have the same symmetry property than the initial Euler
equations.

The aim of our study is to determine an expression of the turbulent
current \( \bf {J_{\omega }} \) in terms of the average quantities.
This closure problem is equivalent to the usual one for the velocity
correlations, when the evolution of the velocity is considered.

\section{Anisotropy of the turbulent current\label{sec:Flux}}

In this section, we first derive an approximation of the turbulent
current, as the leading order of an expansion in power of the homogenization
scale \( l \) (the actual expansion parameter is \( l/L \), where
\( L \) is a typical size of the domain \( D \)). We discuss the
main properties of this deterministic expression. The turbulent current
is anisotropic. It is related to the symmetric part of the strain
tensor for the averaged velocity. It exactly conserves the energy
of the averaged fields. Using typical vorticity fields of freely decaying
turbulence, we numerically compute the actual turbulent current, in
order to study the validity of this first order approximation. \\

\subsection{A leading order deterministic approximation\label{sec_Deterministe}}

In order to obtain an approximation of the turbulent current \( {\bf J}_{\omega } \),
we decompose the vorticity field as the sum of its average \( \overline{\omega } \)
and of fluctuations \( \tilde{\omega } \), \( \tilde{\omega } \)
being defined by \( \omega =\overline{\omega }+\tilde{\omega } \).
Similarly \( {\bf u}=\overline{{\bf u}}+\widetilde{{\bf u}} \). We
then obtain for the turbulent current \( {\bf J}_{\omega } \) (\ref{Euler_moy}):\foreignlanguage{english}{\begin{equation}
\label{Courant_Turbulent_Decomposition_Classique}
{\bf J}_{\omega }\equiv \overline{\omega {\bf u}}-\overline{\omega }\, \overline{{\bf u}}=\underbrace{\underbrace{\left( \overline{\overline{\omega }\, \overline{{\bf u}}}-\overline{\omega }\, \overline{{\bf u}}\right) }_{1}+\underbrace{\overline{\tilde{\omega }\overline{{\bf u}}}}_{2}}_{{\bf J}_{d}}+\underbrace{\underbrace{\overline{\overline{\omega }\widetilde{{\bf u}}}}_{3}+\underbrace{\overline{\tilde{\omega }\widetilde{{\bf u}}}}_{4}}_{{\bf J}_{s}}
\end{equation}
} 

We note that the first term (double average), the second term (correlation
between the vorticity fluctuations and the mean velocity) and the
third term (correlation between the velocity fluctuations and the
mean velocity) are all identically equal to zero for an ensemble average.
The decomposition (\ref{Courant_Turbulent_Decomposition_Classique})
is the commonly used one. However, for reason that shall become clear
we will use :\begin{equation}
\label{Flux_Deterministe+Stochastique}
{\bf J}_{\omega }={\bf J}_{d}+{\bf J}_{s}\, \, \, \, {\rm with}\, \, \, \, {\bf J}_{d}=\overline{\overline{\omega }\overline{{\bf u}}}+\overline{\tilde{\omega }\overline{{\bf u}}}-\overline{\omega }\, \overline{{\bf u}}\, \, \, \, {\rm et}\, \, \, \, {\bf J}_{s}=\overline{\overline{\omega }\widetilde{{\bf u}}}+\overline{\tilde{\omega }\widetilde{{\bf u}}}
\end{equation}
The first term \( {\bf J}_{d} \), which does not depend on the velocity
fluctuations, will be called the deterministic component of the turbulent
current (we will show that it can be expressed, to a good approximation,
in terms of the averaged quantities). The second term depends on the
velocity fluctuations. It will be called the stochastic component
of the turbulent current. As the order of magnitude of \( \widetilde{{\bf u}} \)
is \( l \) times the order of magnitude of \( {\bf u} \), we anticipate
that the stochastic current will be negligible at leading order in
\( l \).

In order to relate the deterministic part of the turbulent current
\( {\bf J}_{d}=\overline{\omega \overline{{\bf u}}}-\overline{\omega }\, \overline{{\bf u}} \)
(\ref{Flux_Deterministe+Stochastique}) to the strain tensor for the
averaged velocity, we expand \( \overline{{\bf u}} \) in Taylor series
around a point \( O \). By noting that the zeroth order terms are
null, we obtain at the leading order in \( l \): \( {\bf J}_{d}\left( O\right) =\overline{\omega x}\left( O\right) \partial _{x}\overline{u_{x}}\left( O\right) +\overline{\omega y}\left( O\right) \partial _{y}\overline{u_{x}}\left( O\right)  \).
Using that the the average operator in equivalent to a Gaussian for
small \( l \), we note that a part-integration leads to \( \overline{\omega x}\left( O\right) =l^{2}\partial _{x}\overline{\omega }\left( O\right)  \)
and \( \overline{\omega y}\left( O\right) =l^{2}\partial _{y}\overline{\omega }(O) \).
Using this result, we obtain that the deterministic part of the turbulent
current is equal, at leading order, to the action of the strain tensor
on the vorticity gradient:\[
{\bf J}_{d}=l^{2}{\bf \Sigma }{\bf \nabla }\overline{\omega }\, \, \, \, {\rm (order\, 1)}\, \, \, \, \, \, {\rm with}\, \, \, \, {\bf \Sigma }=\left( \begin{array}{cc}
\partial _{x}\overline{u_{x}} & \partial _{y}\overline{u_{x}}\\
\partial _{x}\overline{u_{y}} & \partial _{y}\overline{u_{y}}
\end{array}\right) \]
As the typical variation length for the mean vorticity \( \overline{\omega } \)
is \( l \), its gradient is of order \( 1/l \), and this expression
is actually of order 1. This expansion may be led to any order in
\( l \). At each order it involves derivatives of \( \overline{\omega } \)
at larger orders. In section \ref{sec_Courant_Turbulent_Numerique}
we will study numerically this approximation for vorticity fields
typical of freely decaying turbulence, in the regime of vortex merging
and filamentation.

The antisymmetric part of the strain tensor, \( {\bf \Sigma }_{ant} \),
characterizes the local rotation of the fluid. It is linked to the
vorticity by: \( {\bf \Sigma }_{ant}=2\left( \begin{array}{cc}
0 & \overline{\omega }\\
-\overline{\omega } & 0
\end{array}\right)  \). It is easily verified that \( \nabla .\left( {\bf \Sigma }_{ant}{\bf \nabla }\overline{\omega }\right) =0 \).
Thus, whereas it is present in the leading order approximation for
\( {\bf J}_{d} \), this antisymmetric part has no effect on the evolution
of the vorticity. We have: \begin{equation}
\label{Flux_Deterministe_Ordre2}
\nabla .{\bf J}_{d}=l^{2}\nabla .\left( {\bf \Sigma }_{sym}{\bf \nabla }\overline{\omega }\right) \, \, \, \, {\rm (order\, \, 1)}\, \, \, \, \, \, {\rm with}\, \, \, \, {\bf \Sigma }_{sym}=\left( \begin{array}{cc}
\partial _{x}\overline{u_{x}} & \left( \partial _{y}\overline{u_{x}}+\partial _{x}\overline{u_{y}}\right) /2\\
\left( \partial _{y}\overline{u_{x}}+\partial _{x}\overline{u_{y}}\right) /2 & \partial _{y}\overline{u_{y}}
\end{array}\right) 
\end{equation}
The symmetric part of the strain tensor may be diagonalized with orthogonal
eigenvectors. As the fluid is incompressible, it has a null trace
and thus has two opposite eigenvalues \( \sigma  \) et \( -\sigma  \),
with eigenvectors \( {\bf e}_{\sigma } \) and \( {\bf e}_{-\sigma } \)
respectively. \( \sigma  \) is the mean strain: 

\[
\sigma =\sqrt{-{\rm det}\left( {\bf \Sigma }_{sym}\right) }=\sqrt{-\partial _{x}u_{x}\partial _{y}u_{y}+\left( \partial _{y}u_{x}-\partial _{x}u_{y}\right) ^{2}/4}\]

Let us consider the operator \( \overline{\omega }\rightarrow \int _{D}d{\bf r}\, \overline{\omega }\nabla .\left( {\bf \Sigma }_{sym}{\bf \nabla }\overline{\omega }\right)  \).
The contraction direction of the strain tensor \( {\bf e}_{\sigma } \)
corresponds to the definite negative part of this operator: for any
\( \overline{\omega } \), \( \int _{D}d{\bf r}\, \overline{\omega }\nabla .\left( {\bf e}_{\sigma }{\bf e}_{\sigma }.{\bf \nabla }\overline{\omega }\right) <0 \)
(by part integration). It thus acts as a positive viscosity. Conversely,
the stretching direction of the strain \( {\bf e}_{-\sigma } \) tensor
will correspond to the positive part of this operator. It thus act
as a negative viscosity. We will note \( {\bf \Sigma }_{sym}={\bf \Sigma }^{>}_{sym}+{\bf \Sigma }^{<}_{sym} \)
with \( {\bf \Sigma }^{<}_{sym}=\sigma {\bf e}_{\sigma }\, ^{T}{\bf e}_{\sigma } \)
and \( {\bf \Sigma }^{>}_{sym}=-\sigma {\bf e}_{-\sigma }\, ^{T}{\bf e}_{-\sigma } \)
where \( ^{T} \) is the transposition. From a numerical point of
view, the negative part of the symmetric part of the strain tensor
\( {\bf \Sigma }^{<}_{sym} \) will have a stabilization effect whereas
the positive part \( {\bf \Sigma }^{>}_{sym} \) will have a destabilization
effect. In a numerical code, this last term will lead to instabilities.
For this reason, the approximate expression (\ref{Flux_Deterministe_Ordre2})
does not suit for a direct parameterization of the effect of small
scales on large scales.

To conclude this analysis, we note that the leading order approximation
of the turbulent current (\ref{Flux_Deterministe_Ordre2}) exactly
conserves the energy of the averaged field : \( \overline{E}=\frac{1}{2}\int _{D}d{\bf r}\, \overline{{\bf u}}^{2}=\frac{1}{2}\int _{D}d{\bf r}\, \overline{\omega }\, \overline{\psi } \).
The conservation of the energy is a key property of two-dimensional
turbulence. In section \ref{sec_Courant_Turbulent_Numerique}, we
will show numerically that the actual turbulent current (not the leading
order approximation) is responsible for a very small flux of energy
towards the large scale (\( \overline{E} \) slightly increase). This
is in accordance with the phenomenology of two-dimensional freely
decaying turbulence.\\

We have described a leading order approximation for the turbulent
current. This approximation is deterministic (it can be expressed
explicitly in terms of the averaged quantities). The turbulent current
is the application of the strain tensor on the gradient of the mean
vorticity. This approximation exactly conserves the energy. In the
following section we will show numerically that this approximation
is a go one for typical vorticity fields. However, this analytical
expression can not be used to design a stable numerical scheme. In
section \ref{sec : Relaxation_Anisotrope}, we will propose a stable
scheme that respects the main properties of this approximation.

\subsection{Numerical analysis of the turbulent current\label{sec_Courant_Turbulent_Numerique}}

For a given vorticity field, it is possible to compute numerically
the averaged field at a given scale. From it, one can then compute
the turbulent current. The aim of this section is to describe qualitatively
the turbulent current, and to study the validity of the leading order
approximation, for vorticity fields which are typical of freely decaying
turbulence.\\

\begin{figure}
{\centering \vskip-2cm\begin{tabular}{c}
\selectlanguage{french}
\resizebox*{9cm}{9cm}{\includegraphics{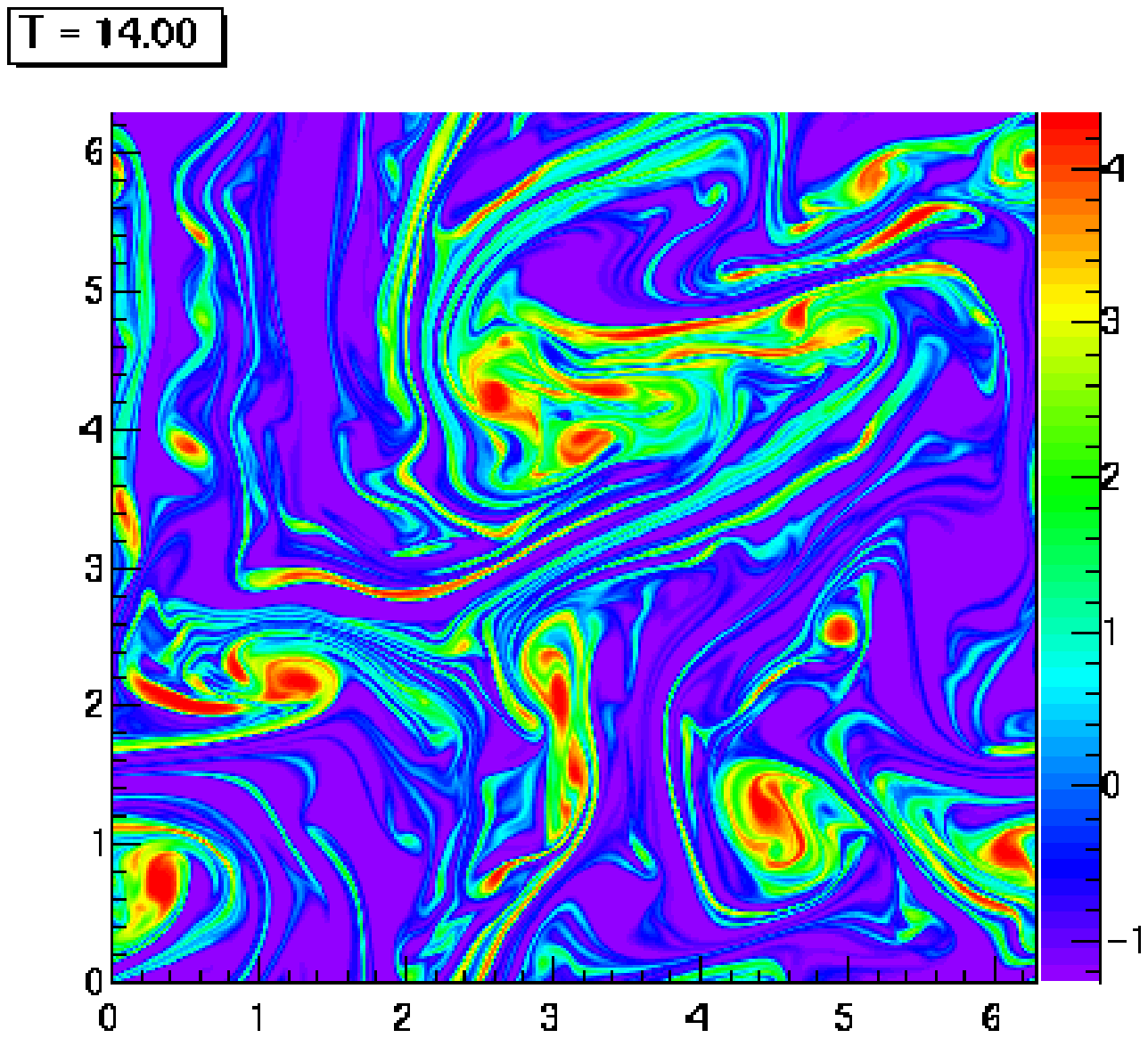}} 
\selectlanguage{english}\\
\selectlanguage{french}
\resizebox*{9cm}{9cm}{\includegraphics{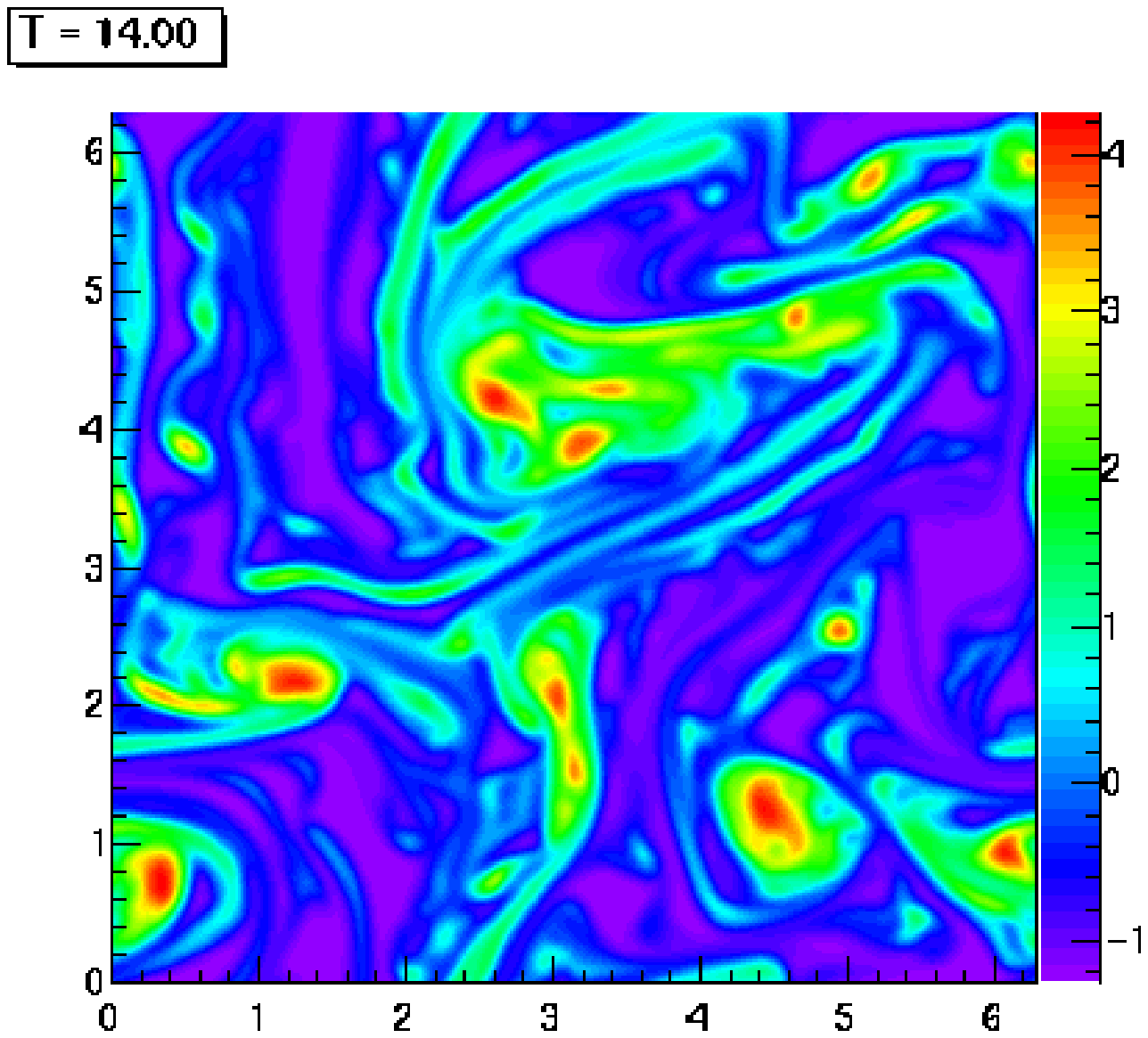}} 
\selectlanguage{english}\\
\end{tabular}\par}

\caption{\label{fig:vp-15}Upper picture: a typical vorticity field \protect\( \omega \protect \)
for freely decaying turbulence in the regime of filamentation and
vortex coalescence (resolution 1024X1024). Lower picture: the corresponding
averaged vorticity field \protect\( \overline{\omega }\protect \),
for a Gaussian homogenization at the scale \protect\( l=2\pi /128\simeq 0.049\protect \).
Fig. \ref{fig:Divergence-Flux-15} shows the corresponding turbulent
current. }
\selectlanguage{english}
\end{figure}

In order to obtain freely decaying turbulence typical vorticity fields,
we use results of numerical simulations. For all numerical computations
of this article, we will use a pseudo-spectral code, with an order
3 Adam-Bradsforth scheme for the temporal discretisation, on a domain
periodic in both direction. The result are presented for a vorticity
field obtained by numerically solving the Navier Stokes equations: 

\begin{equation}
\label{QG_Navier_Stokes}
\frac{\partial \omega }{\partial t}+{\bf u}.\nabla \omega =\nu \Delta \omega 
\end{equation}
For a given resolution, the viscosity \( \nu  \) is chosen as small
as possible, such that the computation is stable. The Navier Stokes
equation verify a maximum principle : the maxima (resp. minima) of
the vorticity must always decrease (resp. increase) in time. In order
to test the stability of the computation, we check for any computation
the extrema of the vorticity. We have studied the turbulent current
for several situations. We discuss the results for the vorticity field
represented on Fig. \ref{fig:vp-15}. It has been obtained by considering
an initial condition made of random patches of potential vorticity.
The typical vorticity and velocity are of order 1. The parameter of
the computation are \( \nu =3.14\, 10^{-5}, \) \( Re=2\pi u_{max}/\nu =200000, \)
\( dt=6.14\, 10^{-4} \) (time step), resolution \( 1024X1024 \).
The vorticity field is the one corresponding to \( t=15 \). At this
time, the decrease of the total energy is of order \( 2\% \), such
that the Navier-Stokes approximation may be considered a good approximation
of the Euler equation. A more precise description of the computation
is given in Bouchet (2001).

This vorticity field is characteristic of the regime of strong filamentation,
due to the nonlinear interaction between the vortex. Fig. \ref{fig:vp-15}
shows that coherent vortices are present together with filament structures,
at all scale of the field. We will compute the averaged vorticity
field, using the averaging operator described in section \ref{sec: moyenis=E9e},
with \( l=2\pi /128. \) This corresponds to an homogenization at
a scale corresponding to a numerical resolution \( 128X128 \). The
corresponding averaged vorticity field is represented on Fig. \ref{fig:vp-15}.\\

\begin{figure}
{\centering \resizebox*{10cm}{10cm}{\includegraphics{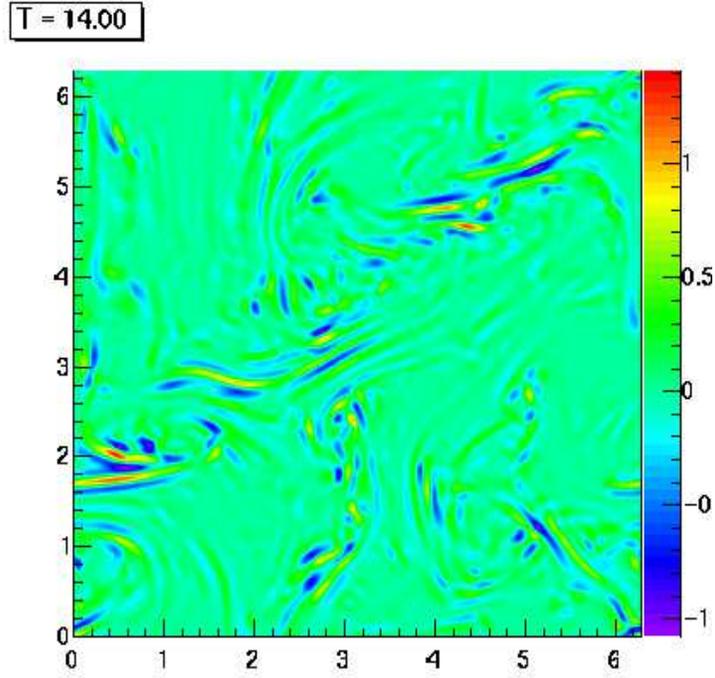}} \par}

\caption{\label{fig:Divergence-Flux-15} The divergence of the turbulent current
\protect\( \nabla .{\bf J}_{\omega }\protect \), computed for a Gaussian
homogenization at a scale \protect\( l=2\pi /128\simeq 0.049\protect \),
for the vorticity field shown on Fig. \ref{fig:vp-15}. The comparison
with Fig. \ref{fig:vp-15} illustrates that important values of the
divergence of the turbulent current are associated with structures
with typical scale \protect\( l\protect \), especially filaments.
Not all structures at scale \protect\( l\protect \) give a strong
contribution, but only the ones correlated with a strong mean strain
(see Fig. \ref{fig:cisaillement-15}).}
\selectlanguage{english}
\end{figure}

\begin{figure}
{\centering \resizebox*{10cm}{10cm}{\includegraphics{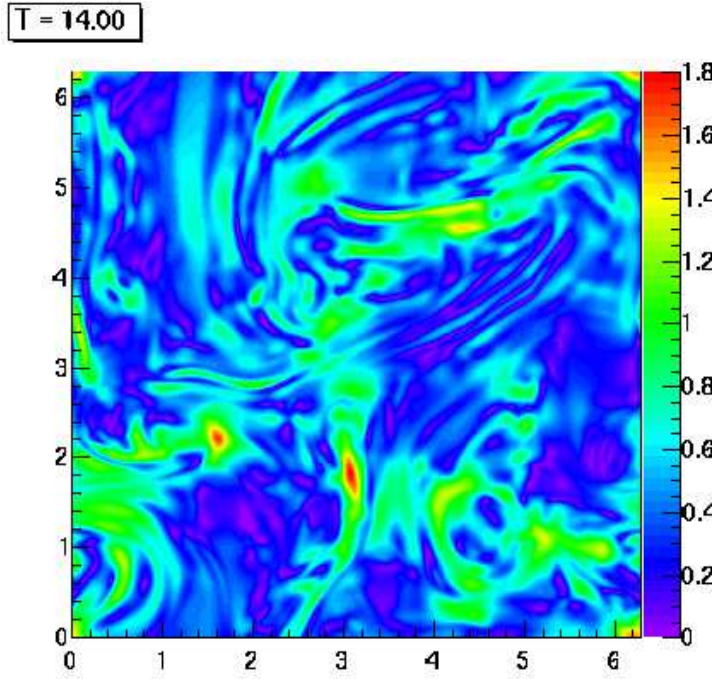}} \par}

\caption{\label{fig:cisaillement-15} The mean strain \protect\( \sigma \protect \)
(positive eigenvalue of the symmetric part of the strain tensor) for
the velocity field corresponding to the homogenized vorticity \protect\( \overline{\omega }\protect \)
(Fig. \ref{fig:vp-15}). We note the correlation between areas of
strong strain, areas with strong vorticity gradients (Fig. \ref{fig:vp-15})
and areas where the divergence of the turbulent current is high (Fig.
\ref{fig:Divergence-Flux-15}). }
\selectlanguage{english}
\end{figure}

From the averaged vorticity field, we compute the turbulent current
\( \bf {J}_{\omega }=\overline{\omega {\bf u}}-\overline{\omega }\, \overline{{\bf u}} \).
Fig. \ref{fig:Divergence-Flux-15} shows the divergence \( \nabla .{\bf J}_{\omega } \)
of the turbulent current, associated to the vorticity field \( \omega  \).
The comparison of this picture with the vorticity one (Fig. \ref{fig:vp-15})
clearly illustrates that the divergence of the turbulent current is
associated to structures with typical scales of the order of the homogenization
scale \( l \). Most of these structures are filaments that are stretched
by the flow strain, passing from resolved to unresolved scales. Fig.
\ref{fig:cisaillement-15} shows the mean strain \( \sigma  \), for
the averaged velocity field. The strain is a vectorial property. This
figure however clearly show that areas with strong strain are associated
with strong vorticity gradients (alignment of the structures with
the strain), and that these areas are the ones that give strong contributions
to the divergence of the turbulent current. This qualitative description
is in accordance with the result we have obtained for the leading
order approximation of the turbulent current by the effect of the
strain tensor on the gradient of the vorticity field.

Let us analyze quantitatively the contribution of the various terms
composing the turbulent current. Table \ref{tab:flux} shows the divergence
of the terms of the classical decomposition (\ref{Courant_Turbulent_Decomposition_Classique}).
This illustrates that the two first terms have a contribution much
larger than the total turbulent current. We thus note that these two
first current are strongly anti-correlated. Indeed their sum \( {\bf J}_{d} \)
give a contribution much less important that each of its components.
Table \ref{tab:flux} also shows that the principal contribution of
\( {\bf J}_{s} \) is the correlation between vorticity fluctuations
and velocity fluctuations. In a numerical simulation with resolved
scales up to the scale \( l \), this will correspond to the contribution
of unresolved scales, for which any information is loss. A statistical
modeling of this component will then be the more natural. That's why
we call \( {\bf J}_{s} \), the stochastic component of the turbulent
current. The deterministic part dominates the stochastic part, by
a factor 10 to 15, for the three studied norms. This is in accordance
with the leading order approximation we have obtained in section \ref{sec_Deterministe}.

Let us discuss the results obtained by Laval, Dubrulle and Nazarenko
\cite{Laval_Dubrulle_Nazarenko_99}, for currents computed for an
average corresponding to a spectral truncation of the vorticity field.
They have studied the relative contribution of the currents 2, 3 et
4 (see \ref{Courant_Turbulent_Decomposition_Classique}). They have
shown that the advection of the fluctuations by the average velocity
(term 2) dominate the terms 3 and 4. This is in accordance with the
result we have obtained with a Gaussian homogenization. Using these
results, they have justified an approximation neglecting these last
two terms. They have then proposed a parameterization of small scale
effects, in which the second term is determined thanks to a particle
description of the fluctuations \cite{Laval_DN_Phys_Fluids}. We will
follow another approach: we use a deterministic approximation of the
sum of the two terms 1+2=\( {\bf J}_{d} \).

Table \ref{tab:flux_cisaillement} shows the norms of the difference
between \( {\bf J}_{d} \) and its first (\ref{Flux_Deterministe_Ordre2})
and second order approximation, in the small \( l/L \) expansion.
We conclude that, for this vorticity field, in the regime of vortex
merging and strong filamentation, the deterministic part of the turbulent
current can be approximated by the action of the strain tensor on
the gradient of the vorticity, up to an error of order of 10\%. Table
\ref{tab:flux_cisaillement} also shows the relative contribution
of the positive and of the negative parts of the strain tensor. The
negative part (\( {\bf \Sigma }^{<}_{sym} \), positive eigenvalue)
has a larger contribution than the positive ones (\( {\bf \Sigma }^{>}_{sym} \),
negative eigenvalue). This illustrates the irreversibility of the
flow evolution: due to this property, the moments of the vorticity
will evolved towards values corresponding to the further mixing of
the vorticity. We however recall that energy loss associated to the
negative part is exactly compensated by energy gain associated to
the positive part.\\

\section{Anisotropic relaxation equations\label{sec : Relaxation_Anisotrope} }

In section \ref{sec:Flux} we have obtained a deterministic approximation
for the turbulent current \( \bf {J}_{\omega } \). This approximation
is the action of the strain tensor on the gradient of the vorticity.
It exactly conserves the energy. However the positive part of the
strain tensor acts similarly to a negative viscosity. This would render
any numerical simulation unstable. In this section, in order to obtain
a parameterization as close as possible to the actual turbulent current,
but with the essential practical constraint to lead to a stable numerical
scheme, we will search for a turbulent current which have the property
of the deterministic one in the contraction direction of the strain
tensor (stabilization effect), and which exactly conserves the energy. 

There are several ways to impose such constraints to a parameterization.
We choose to determine a parameterization which is in accordance with
the statistical properties of the mixing of the vorticity. Equilibrium
statistical mechanics of the two-dimensional Euler equations describes
the most probable mixing of the vorticity for a given distribution
of potential vorticity and Energy. We will then choose the turbulent
current which maximize the entropy production, while respecting the
energy conservation and the effect of the stable contribution of the
strain tensor. 

A similar maximum entropy production principle (MEPP) has been used
by Robert and Sommeria \cite{Robert_Sommeria_92} in order to parameterize
the small scale turbulence. In this work, author however impose an
isotropic constraint on the turbulent current. We make a derivation
very close to their one, but we impose the value of the turbulent
current in the stable direction of the strain tensor. We thus explicitly
take into account the anisotropic nature of the turbulent current.
We do not think that such a maximum entropy production principle is
based on any physical or theoretical ground. The anisotropic relaxation
turbulent current will indeed be different from the observed one.
We however use this principle to obtain stable equation as close as
possible from the correct physical ones. 

In section \ref{sec: Comparaison_numerique}, we will show that the
obtained anisotropic relaxation equations lead to better results than
classical parameterizations in order to determine the evolution of
coarse\_grained flow.\\

Let us first briefly introduce the results of the equilibrium statistical
mechanics of the vorticity. In order to simplify this discussion,
we will consider an initial condition made of many vorticity patches,
but with only two values \( \omega =a_{1} \) and \( \omega =a_{2} \),
occupying respectively areas \( A_{1} \) and \( A_{2} \). Generalization
of this discussion and of the following results to any initial distribution
of vorticity is however straightforward. We note that the initial
distribution (values \( a_{1} \) and \( a_{2} \), and areas \( A_{1} \)
and \( A_{2} \)) are conserved by the inviscid dynamics. We now consider
a coarse-graining of the vorticity at a scale \( l \) (local-average
at the scale \( l \), which will be interpreted in our case, as the
numerical resolution scale). At this coarse-grained scale (macroscopic
description), we will describe the vorticity fields by the local probabilities
\( p({\bf r}) \) and \( (1-p({\bf r})) \) to have respectively the
vorticity values \( a_{1} \) and \( a_{2} \), at the position \( {\bf r} \).
The expression of the coarse-grained vorticity is then: \( \overline{\omega }=a_{1}p+a_{2}(1-p) \).
The main result of the equilibrium statistical mechanics, is the evaluation
of the probability to observe a local probability field \( p({\bf r}) \),
given the vorticity distribution a given energy \( E \). When the
scale \( l \) goes to zero, the logarithm of the probability of two
probability fields is given by the entropy \( S \):

\begin{equation}
\label{Entropie}
S=\int _{D}d{\bf r}\, s(p)\, \, \, \, {\rm with}\, \, \, \, s(p)=-\left( p\log p+(1-p)\log (1-p)\right) 
\end{equation}
(see \cite{Miller,Robert_Sommeria_91}, and \cite{Michel_Robert_Large_Deviation,Robert_2000}
for a justification). The most probable state is then given by the
maximization of the entropy for a given energy and vorticity distribution.
This equilibrium is a stationary state of the dynamical equations.
The main hypothesis of the equilibrium statistical mechanics is that
the complex dynamical evolution of the flow will lead to a state close
to the equilibrium one.\\

We would like to describe the coarse grained evolution of the vorticity.
On the contrary to the equilibrium case, no clear principle permits
to describe this out of equilibrium relaxation. The Euler equations
advect the two levels of vorticity with the velocity \( {\bf u} \).
This conservation equation will lead to a transport equation for \( p \),
by the mean velocity \( \overline{{\bf u}} \) on one hand, and by
a turbulent current \( {\bf {\cal J}} \) as a consequence of the
unresolved scales:\begin{equation}
\label{Evolution_p}
\partial _{t}p+\overline{{\bf u}}.\nabla p=-\nabla .{\bf {\cal J}}
\end{equation}
Using the link between the averaged vorticity and the probability:
\( \overline{\omega }=a_{1}p+a_{2}(1-p) \), the coarse-grained vorticity
verify equation (\ref{Euler_moy}) with \( {\bf J}_{\omega }=(a_{1}-a_{2}){\bf {\cal J}} \).
Our aim is thus to determine \( \bf {J}_{\omega } \). \\

As explained in the introduction, we look for a turbulent current
which has a determined behavior in a direction given by the strain
tensor and which conserve the energy. As this does not specify the
turbulent current, in order to obtain an equation compatible with
the statistical tendency of the equation to increase the entropy,
we will look at equation which maximize the entropy production rate,
given these dynamical constraints.

Using (\ref{Evolution_p}), we first compute the entropy and energy
production rates:\begin{equation}
\label{Variation_entropie}
\frac{dS}{dt}=-\int _{D}d{\bf r}\, \frac{\nabla \overline{\omega }.{\bf J}_{\omega }}{\left( a_{1}-\overline{\omega }\right) \left( \overline{\omega }-a_{2}\right) }
\end{equation}

\begin{equation}
\label{Variation_energie}
\rm {and}\, \, \, \, \frac{dE}{dt}=\int _{D}d{\bf r}\, \nabla \psi .{\bf J}_{\omega }
\end{equation}
 We note that \( \left( a_{1}-\overline{\omega }\right) \left( \overline{\omega }-a_{2}\right) \geq 0 \).
This expression shows that the entropy production is maximal when
the current has a direction opposite to the one of the vorticity gradient. 

We want to impose an anisotropic constraint on the turbulent current
\( \bf {J}_{\omega } \) ; we put: \begin{equation}
\label{Energie_diffusion_anisotrope}
C_{1}({\bf r})\frac{\left( {\bf J}_{\omega }.{\bf e}_{1}({\bf r})\right) ^{2}}{2}+C_{2}({\bf r})\frac{\left( {\bf J}_{\omega }.{\bf e}_{2}({\bf r})\right) ^{2}}{2}=1
\end{equation}
 where the two vectors \( {\bf e}_{1} \) and \( {\bf e}_{2} \) (depending
on \( {\bf r} \)), and the values of \( C_{1} \) and \( C_{2} \)
will be specified latter. We search for the current \( \bf {J}_{\omega } \)
which maximize the entropy production (\ref{Variation_entropie}),
while verifying the energy conservation (\ref{Variation_energie}),
the anisotropic constraint (\ref{Energie_diffusion_anisotrope}) and
the global constraint on \( \bf {J}_{\omega } \): \( \int _{D}d{\bf r}\, {\bf J}_{\omega }=0 \).
We thus consider the critical points of the functional: \[
\frac{dS}{dt}-\beta \frac{dE}{dt}-\sum _{i=1,2}\int _{D}d{\bf r}\, \alpha ({\bf r})C_{i}({\bf r})\left( {\bf J}_{\omega }.{\bf e}_{i}\right) ^{2}+\left( {\bf e}_{z}\wedge {\bf V}\right) .\int _{D}d{\bf r}\, {\bf J}_{\omega },\]
where \( \beta  \), \( \alpha ({\bf r}) \) and \( -{\bf e}_{z}\wedge {\bf V} \)
are Lagrange parameters associated to the constraints. The first variations
of this functional lead to:\begin{equation}
\label{courant_MEPP}
{\bf J}_{\omega }=-\sum _{i=1,2}\nu _{i}\left[ \left( {\bf \nabla }\overline{\omega }+\left( a_{1}-\omega \right) \left( \omega -a_{2}\right) \left( \beta {\bf \nabla }\overline{\psi }-{\bf e}_{z}\wedge {\bf V}\right) \right) .{\bf e}_{i}\right] \, {\bf e}_{i}
\end{equation}
with \( \nu _{i}=-1/\left( \alpha C_{i}\left( a_{1}-\overline{\omega }\right) \left( \overline{\omega }-a_{2}\right) \right)  \). 

Using the energy constraint (\ref{Variation_energie}) and the condition
on the zero global average for \( {\bf J}_{\omega } \), we compute
the entropy production:

\[
\frac{dS}{dt}=\int _{D}d{\bf r}\, \sum _{i=1,2}\frac{\left( {\bf J}_{\omega }.{\bf e}_{i}\right) ^{2}}{\nu _{i}\left( a_{1}-\overline{\omega }\right) \left( \overline{\omega }-a_{2}\right) }\]
Noting that \( \left( a_{1}-\overline{\omega }\right) \left( \overline{\omega }-a_{2}\right) >0 \),
we conclude that if the diffusivities \( \nu _{i} \) are strictly
positive, the entropy variation is actually an entropy production.
With strictly positive diffusivities, we can also conclude, that if
the flow converges towards a stationary state, the entropy production
must vanish and thus the current \( {\bf J}_{\omega } \) must be
zero. In such a case, from (\ref{courant_MEPP}) we obtain \( {\bf e}_{z}\wedge {\bf V}={\bf \nabla }\left( \beta \left( a_{1}-a_{2}\right) \overline{\psi }+\log \left( \overline{\omega }-a_{2}\right) -\log \left( a_{1}-\overline{\omega }\right) \right)  \).
The constant \( {\bf e}_{z}\wedge {\bf V} \) being a gradient must
be equal to zero. This equation can then be integrated to. This yields:
\begin{equation}
\label{Etat_Gibbs_Euler}
\overline{\omega }=-\Delta \overline{\psi }=\frac{a_{1}+a_{2}}{2}-\tanh \left( \frac{a_{1}-a_{2}}{2}\left( \beta \overline{\psi }+\alpha \right) \right) 
\end{equation}
This is the equations of the statistical equilibrium, which is a stationary
state of the Euler equations. We thus conclude, that if the equations
for the coarse-grained field, with turbulent current (\ref{courant_MEPP}),
converge ; they converge towards a statistical equilibrium. We will
call these equation anisotropic relaxation equations.\\

Let us determine \( \beta  \) and \( \bf {V} \). For this let us
denote \( {\bf A}\equiv \sum _{i=1,2}\int _{D}d{\bf r}\, \nu _{i}\left( a_{1}-\overline{\omega }\right) \left( \overline{\omega }-a_{2}\right) {\bf e}i\, ^{T}{\bf e}_{i} \),
(\( ^{T} \) is the transposition operator). \( {\bf A} \) is thus
an order two symmetric tensor), \( {\bf b}=\sum _{i=1,2}\int _{D}d{\bf r}\, \nu _{i}\left( a_{1}-\overline{\omega }\right) \left( \overline{\omega }-a_{2}\right) \left( \nabla \psi .{\bf e}_{i}\right) {\bf e}_{i} \)
(2d vector) and \( c=\sum _{i=1,2}\int _{D}d{\bf r}\, \nu _{i}\left( a_{1}-\overline{\omega }\right) \left( \overline{\omega }-a_{2}\right) \left( \nabla \overline{\psi }.{\bf e}_{i}\right) ^{2} \)
(scalar). \( M=\left( \begin{array}{cc}
\bf {A} & -\bf {b}\\
-^{t}\bf {b} & c
\end{array}\right)  \) is a 3X3 symmetric matrix. We then express the energy conservation
(\ref{Variation_energie}) and the property of the zero average for
\( {\bf J}_{\omega } \): \( \int _{D}d{\bf r}\, {\bf J}_{\omega }=0 \).
This yields three equations to determine the three variables \( \beta  \)
and \( \bf {V} \):\begin{equation}
\label{M}
\begin{array}{cccc}
M & \left( \begin{array}{c}
{\bf e}_{z}\wedge \bf {V}\\
\beta 
\end{array}\right)  & = & \left( \begin{array}{c}
\sum _{i=1,2}\int _{D}d{\bf r}\, \nu _{i}\left( \nabla \overline{\omega }.{\bf e}_{i}\right) {\bf e}_{i}\\
-\sum _{i=1,2}\int _{D}d{\bf r}\, \nu _{i}\left( \nabla \overline{\omega }.{\bf e}_{i}\right) \left( \nabla \overline{\psi }.{\bf e}_{i}\right) 
\end{array}\right) 
\end{array}
\end{equation}
 We prove in note~%
\footnote{Let us prove that the symmetric matrix \( M \) is invertible. For
this, let us consider \( M \) as defining a quadratic form, and let
us prove that it is definite positive. Using the expressions for \( \bf {A} \),
\( \bf {b} \) and \( c \), the linearity of the integral and the
relation: \( ^{t}{\bf y}{\bf e}_{i}^{t}{\bf e}_{i}{\bf y}=\left( {\bf e}_{i}.{\bf y}\right) ^{2} \)
, a straightforward computation leads to: \( \left( ^{t}{\bf y}\, x\right) M\left( \begin{array}{c}
{\bf y}\\
x
\end{array}\right) =\int _{D}d{\bf r}\, \sum _{i=1,2}\nu _{i}\left( a_{1}-\omega \right) \left( \omega -a_{2}\right) \left( {\bf e}_{i}.\left( \bf {y}-x\nabla \psi \right) \right) ^{2} \) . Given that \( \left( a_{1}-\omega \right) \left( \omega -a_{2}\right) >0, \)
this computation shows that the quadratic form is positive (\( \geq 0) \).
Moreover, noting that \( {\bf e}_{1} \) and \( {\bf e}_{2} \) are
orthonormal, we conclude that if \( \nu _{1} \) and \( \nu _{2} \)
are strictly positives, \( \left( ^{t}{\bf y}\, x\right) M\left( \begin{array}{c}
\bf {y}\\
x
\end{array}\right)  \) can be zero only if \( \left( \bf {y}-x\bf {\nabla \psi }\right)  \)
is null on the whole domain. This would imply that \( \nabla \psi  \)
be constant on the domain. With periodic boundary conditions, this
is possible only if \( \psi =0, \) that is for a flow without motion. 
\selectlanguage{english}
}  that, for strictly positive \( \nu _{1} \) et \( \nu _{2} \),
\( M \) is invertible, except for a flow with \( \psi =0 \), that
is except for an identically zero velocity. \\

We have obtained relaxation equations, which have the property to
increase the entropy while conserving energy. We will now specify
the vectors \( {\bf e}_{i} \) and the values of the diffusivity \( \nu _{i} \).
For this we impose a diffusivity \( l^{2}\sigma  \) in the stable
direction of the strain tensor \( {\bf e}_{\sigma } \) and zero in
the unstable direction of the strain tensor \( {\bf e}_{\sigma } \).
This choice is the one which allow to best approximate the deterministic
approximation of the turbulent current, obtained in section \ref{sec_Deterministe},
while insuring that the diffusivities are positive. We then obtain
the following parameterization:

\begin{equation}
\label{Relaxation_Deformation}
\frac{\partial \overline{\omega }}{\partial t}+\overline{{\bf u}}.\nabla \overline{\omega }=\nabla .\left( l^{2}\sigma \left[ \left( \nabla \overline{\omega }+\beta \left( a_{1}-\overline{\omega }\right) \left( \overline{\omega }-a_{2}\right) \nabla \psi \right) .{\bf e}_{\sigma }\right] {\bf e}_{\sigma }\right) 
\end{equation}
with \begin{equation}
\label{Beta_Relaxation_Deformation}
\beta =-\frac{\int _{D}d{\bf r}\, \sigma \nabla \overline{\omega }.{\bf e}_{\sigma }\nabla \overline{\psi }.{\bf e}_{\sigma }}{\int _{D}d{\bf r}\, \sigma \left( a_{1}-\overline{\omega }\right) \left( \overline{\omega }-a_{2}\right) \left( \nabla \overline{\psi }.{\bf e}_{\sigma }\right) ^{2}}
\end{equation}
To obtain this equation for \( \beta  \) (\ref{Beta_Relaxation_Deformation}),
we have used (\ref{M}) and neglect the effect of the Lagrange parameter
\( \bf {V} \) (see \ref{courant_MEPP}). We have thus assumed \( {\bf V}=0 \).
(in section \ref{sec: Comparaison_numerique}, we will use the relaxation
equation to model flow evolutions. we have done computation with \( \bf {V}\neq 0 \),
using (\ref{M}), but without noting any important difference with
the case \( \bf {V}=0 \)) 

The main hypothesis of the model (\ref{Relaxation_Deformation}) is
that the local, destabilizing action of the positive part of the strain
tensor be replaced by the non local non destabilizing drift term:
\( \beta \left( a_{1}-\overline{\omega }\right) \left( \overline{\omega }-a_{2}\right) \nabla \overline{\psi } \).
This terms insures the energy conservation. \\

From a numerical point of view, the implementation of this equation
has the same level of complexity as the implementation of the Navier-Stokes
equation. All what is further needed is the diagonalization of a 2X2
symmetric matrix to determine the strain tensor directions and the
mean strain \( \sigma  \). The computation cost of these operations
(of order \( N \), where \( N \) is the number of degrees of freedom)
is negligible with respect to the computation of the fast Fourier
transform needed to evaluate the nonlinear terms in a pseudo-spectral
code (of order \( N\log N \)).

In next section, we will compare this model with usual parameterizations
of two-dimensional and geophysical flow computations. We will show
that they give more precise results for the same resolution (or equivalently
for the same computational time). This will characterize an important
predictability gain for the determination of the flow evolution.

\section{Comparison of the anisotropic relaxation equations with commonly
used models \label{sec: Comparaison_numerique}}

The aim of this chapter is to determine the efficiency of the anisotropic
relaxation equations when compared with other models of small scale
turbulence. We consider the regime of vortex merging and filamentation.
We study two types of flows: firstly the merging of four vortices,
which leads to a stable dynamics (the position of the final vortex
is always on the center of the domain) and secondly the merging of
50 vorticity patches which leads to successive vortex merging and
for which the positions of the vortices is very sensitive to the initial
parameterization. In order to compare different models, we will use
as a reference, high resolution direct numerical computations of the
flow. 

We will compare the anisotropic relaxation equation with the following
parameterizations:

\begin{itemize}
\item direct numerical simulations with the Navier Stokes equations (\ref{QG_Navier_Stokes})
(denoted DNS)
\item with hyperviscous parameterization, where the Laplacian operator is
replaced by a bi-Laplacian :\begin{equation}
\label{QG_HV}
\frac{\partial \omega }{\partial t}+{\bf u}.\nabla \omega =\nu \Delta \Delta \omega 
\end{equation}
(denoted HV). Such an hyperviscous parameterization is commonly used
in two-dimensional and geophysical flow computation. It is not based
on any physical ground. However, the bi-Laplacian has the effect to
stabilize the computation whereas it actually acts only on very small
scales (higher effective Reynolds number). It usually dissipates few
energy. It leads to some unphysical spurious description of the small
scale structures, due to the non-respect of the maximum principle
of the Navier Stokes equation (the maxima of the vorticity may increase
in the hyperviscous case)
\item the Smagorinsky model \cite{Smagorinsky} (denoted SMA). The Smagorinsky
model uses an isotropic viscosity \( \nabla .\left( \nu \nabla \omega \right)  \)
but with a non-constant viscosity. The viscosity is taken proportional
to the mean shear \( \sigma  \): \( \nu \left( {\bf r}\right) =l^{2}\sigma \left( {\bf r}\right)  \).
This parameterization takes into account of the importance of the
strain for the transport to small scales of the flow structures. However
the diffusion acts indistinctly in all directions (isotropic). This
model is commonly used in geophysical fluid computation.
\end{itemize}
Concerning the anisotropic relaxation equation, we will alternatively
use either the equations (\ref{Relaxation_Deformation}) and (\ref{Beta_Relaxation_Deformation})
(denoted RELANI) or the equations (\ref{Relaxation_Deformation})
with a local computation of the Lagrange parameter \( \beta  \) for
the energy conservation (denoted RELANILOC). In this last case we
will determine \( \beta  \) with (\ref{Relaxation_Deformation}),
but with an evaluation of the integrals in subdomains \( D_{i} \)
large compared to the resolution scale \( l \), but smaller than
the domain scale. For instance, for low resolution computation with
256X256 degrees of freedom, we will evaluate \( \beta  \) in 12X12
subdomains. We will also consider the isotropic relaxation equations
(Robert and Sommeria \cite{Robert_Sommeria_92}) (denoted REL)

For all of these models, the diffusivities \( \nu  \), or the parameters
\( l \), are always chosen as small as possible for the numerical
computation to be stable.

We quantify the error of the low resolution computation, with respect
to reference high-resolution computation (field denoted by the subscript
\( _{ref} \)), by evaluating the velocity relative error: \( E_{r}\equiv \int _{D}d{\bf r}\, \left( {\bf u}-{\bf u}_{ref}\right) ^{2}/\int _{D}d{\bf r}\, {\bf u}_{ref}^{2} \)
or the vorticity relative error: \( E_{r-\omega }\equiv \int _{D}d{\bf r}\, \left( \omega -\omega _{ref}\right) ^{2}/\int _{D}d{\bf r}\, \omega _{ref}^{2} \).
(these quantities are evaluated in the Fourier space, using a number
of components corresponding to the low resolution fields). \\

\begin{figure*}
{\centering \vskip-20mm\resizebox*{1\textwidth}{1\textheight}{\includegraphics{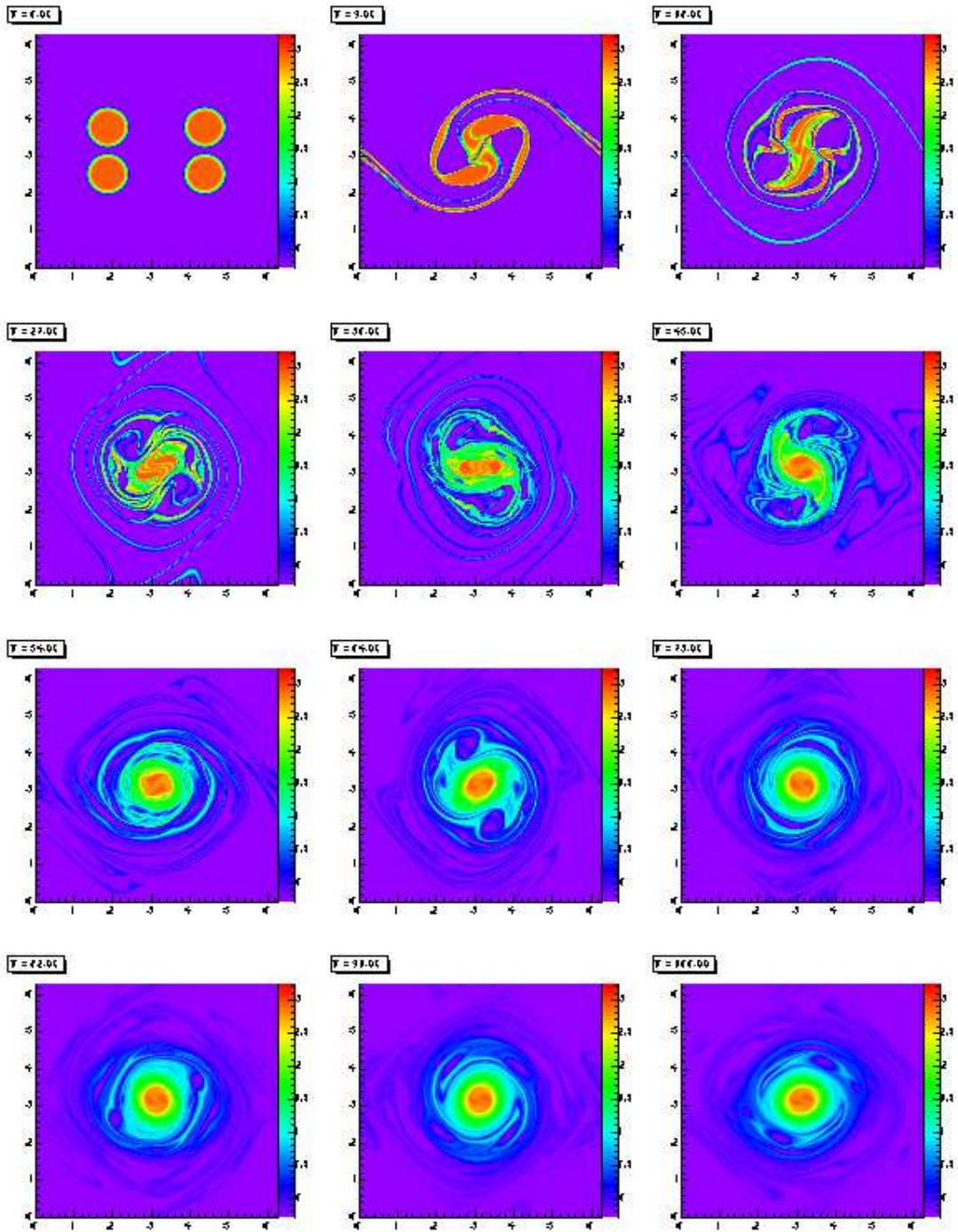}} \par}

\caption{\label{fig:DNSEuler41024-1}Vorticity evolution for the coalescence
of four vortices. This computation uses a Navier-Stokes model, with
resolution 1024x1024 (DNS.Euler4.1024). Time goes from \protect\( T=0\protect \)
to \protect\( T=99\protect \). We consider this computation as representative
of the inertial limit (energy loss of 0.7\% between \protect\( T=0\protect \)
and \protect\( T=100\protect \)). }
\selectlanguage{english}
\end{figure*}

We first consider an initial condition made of four vortices (Fig.
\ref{fig:DNSEuler41024-1}). Initially four vorticity patches with
vorticity \( a_{1}=2.942 \) are in an ambient vorticity \( a_{2}=-0.263 \).
The edges of the vortex patches are smoothed to avoid numerical problems
for initial times. The position of each vortex is slightly moved,
in order to break the apparent symmetry of the initial condition.
We will use as reference computation a Navier-Stokes simulation with
resolution 1024x1024 (DNS.Euler4.1024). Numerical parameters are given
in table \ref{table:parametres_Euler50}. The corresponding Reynolds
number, based on the domain scale, is then \( 400\, 000 \). Fig.
\ref{fig:DNSEuler41024-1} shows the vorticity field evolution. This
typical experiment of vortex merging show the filamentation followed
by a very slow organization towards a stationary state. The energy
loss between times \( T=0 \) and \( T=100 \) is 0.7\%. We thus consider
this computation as close to the inertial limit.

\begin{figure}
{\centering \resizebox*{1\textwidth}{0.5\textheight}{\includegraphics{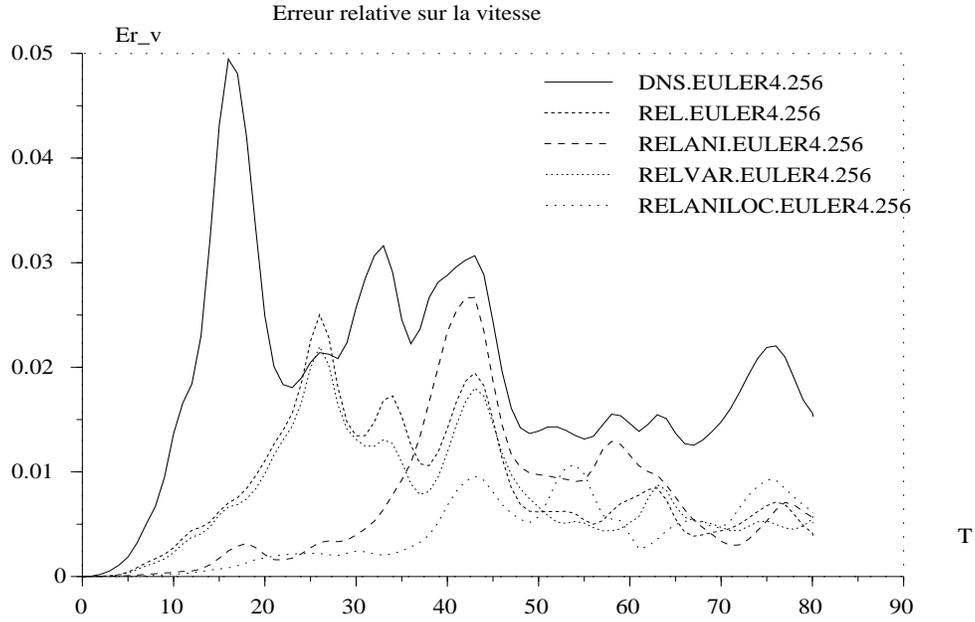}} \par}

\caption{\label{fig: Compare.Euler4.rel.v}Velocity relative error on the
low resolution computation, for the coalescence of four vortices,
for parameterizations of small scales with respectively: a constant
diffusivity (Navier-Stokes) (DNS.Euler4.256), isotropic relaxation
equations with constant diffusion (REL.Euler4.256), anisotropic relaxation
equations (RELANI.Euler4.256) and anisotropic relaxation equations
with local energy conservation (RELANILOC.Euler4.256). }
\selectlanguage{english}
\end{figure}

Fig. \ref{fig: Compare.Euler4.rel.v} shows the evolution of the velocity
relative error for 4 computations with resolutions 256X256 and with
respective parameterizations: Navier Stokes (DNS.Euler4.256), isotropic
relaxation equation (constant diffusivities) (REL.Euler4.256), anisotropic
relaxation equations (RELANI.Euler4.256) and anisotropic relaxation
with local conservation of the energy (RELANILOC.Euler4.256). Table
\ref{table:parametres_Euler50} gives the numerical parameters for
each of these computations. Fig. \ref{fig: Compare.Euler4.rel.v}
shows that the maximal relative error is 5\%. In this case, the numerical
evolution is thus easily predictable. After a quasi-exponential growth,
the error saturates and oscillates together with the position the
oscillation of the slightly asymmetric vortex. The oscillations reflect
the oscillation with respect to the one of the vortex in the reference
computation. The anisotropic relaxation equation with local conservation
of the energy give clearly the best results.

\begin{figure}
{\centering \resizebox*{1\textwidth}{0.5\textheight}{\includegraphics{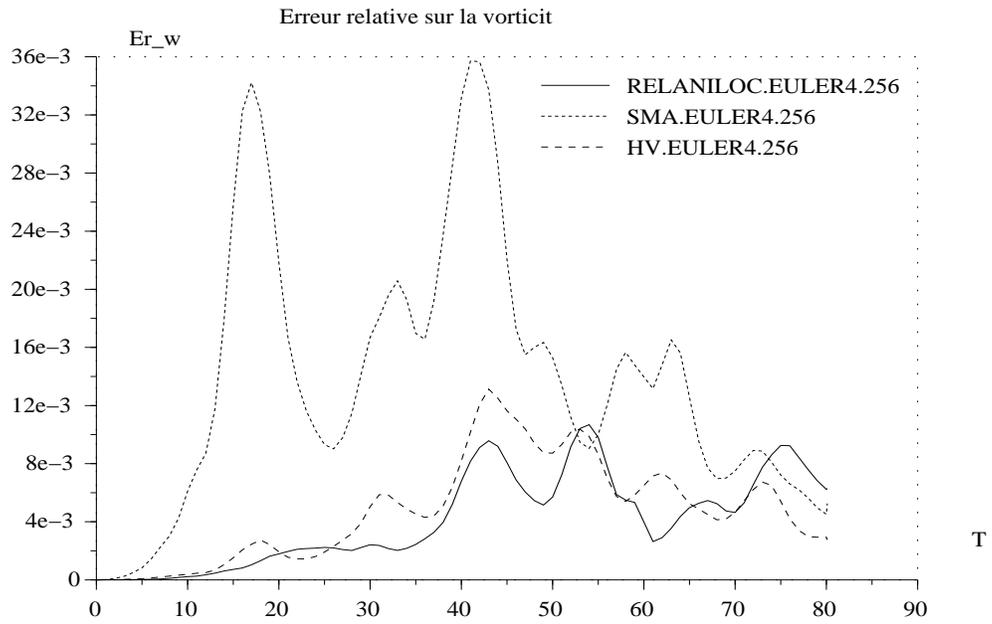}} \par}

\caption{\label{fig: Compare.Euler4.HV.v} Same as Fig. \ref{fig: Compare.Euler4.rel.v},
but for the Smagorinsky model (SMA.Euler4.256), the anisotropic relaxation
equation (RELANI.Euler4.256) and an hyperviscous parameterization
(HV.Euler.4.256) ; for computations with resolution 256x256.}
\selectlanguage{english}
\end{figure}

Fig. \ref{fig: Compare.Euler4.HV.v} compares the evolution of the
vorticity relative error for the anisotropic relaxation equations
with local energy conservation (RELANILOC.Euler4.256), hyperviscosity
(HV.Euler4.256), and the Smagorinsky model (SMA.Euler4.256). For short
times, the anisotropic relaxations give a slightly better result than
the hyperviscous computation. However for larger times, the two model
are not better one than the other. Both models are however much more
efficient than the Smagorinsky model. In \cite{Bouchet_these}, we
present a more complete study and show that same conclusions are obtained
for the study of the evolution of the velocity relative error. For
a smaller resolution 128X128, the anisotropic relaxation equations
turn out to give better results than the hyperviscous parameterization.
\\

\begin{figure*}
{\centering \vskip-20mm\resizebox*{1\textwidth}{1\textheight}{\includegraphics{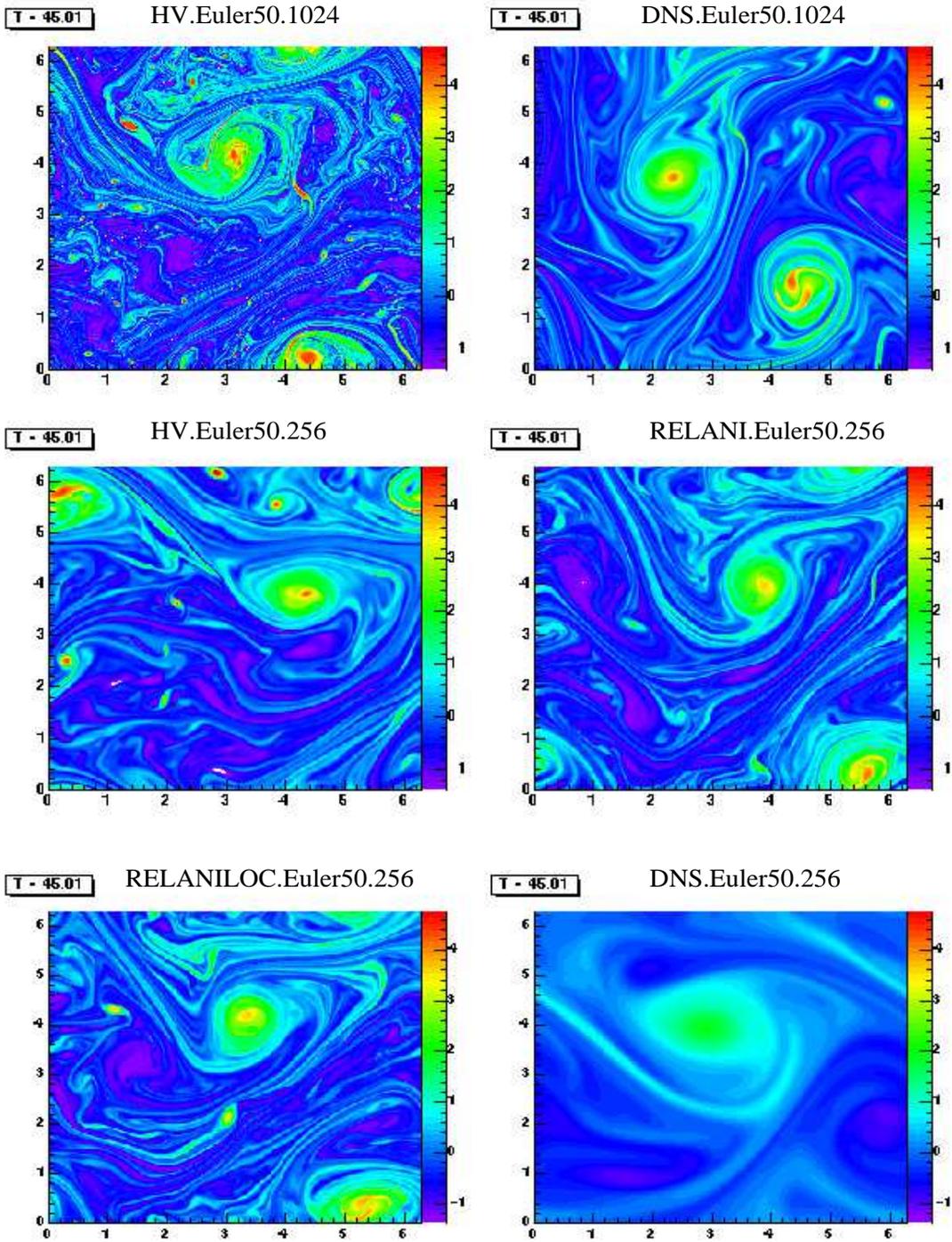}} \par}

\caption{\label{Compare_Plot_Euler50_2}Comparison of vorticity field, obtained
at time \protect\( T=45.01\protect \), from the same initial condition,
for two different high resolution computations: hyperviscous (HV.Euler50.1024)
and Navier Stokes (DNS.Euler50.1024) ; and four low resolution computations
with different small scale turbulence parameterization: hyperviscosity
(HV.Euler50.256), anisotropic relaxation equations (RELANI.Euler50.256),
anisotropic relaxation equations with local enrgy conservation (RELANILOC.Euler50.256)
and Navier Stokes (DNS.Euler50.256).}
\selectlanguage{english}
\end{figure*}

The previous computation is a typical merging of a small number of
vortices. The initial configuration of these vortices leads to a rapid
stabilization of a single structure. In order to study the effect
of the parameterizations for a more complex dynamical evolution, we
will consider an initial vorticity field composed by 50 vorticity
patches, with random initial positions. We make the same study as
in the previous case, comparing the efficiency of the different parameterizations
in a low resolution computation (256X256), using as a reference a
larger resolution computation (1024X1024). As in the previous case,
the Navier-Stokes or Smagorinsky parameterizations lead to a much
less efficient computation than the hyperviscous or anisotropic relaxation
equations ones. For instance, Fig. \ref{Compare_Plot_Euler50_2},
for time \( T=45 \), clearly shows that the Navier-Stokes equation
do not even reproduce the qualitative properties of the vorticity
fields for large time. This is due to the spurious effect of the viscous
dissipation. The Smagorinsky model also leads to a too important diffusion.
(As in previous computations, we alway use the smaller viscosity or
parameter \( l \), compatible with the stability of the computation.
Numerical parameters are reproduced on table \ref{table:parametres_Euler50}.) 

\begin{figure}
{\centering \resizebox*{1\textwidth}{0.5\textheight}{\includegraphics{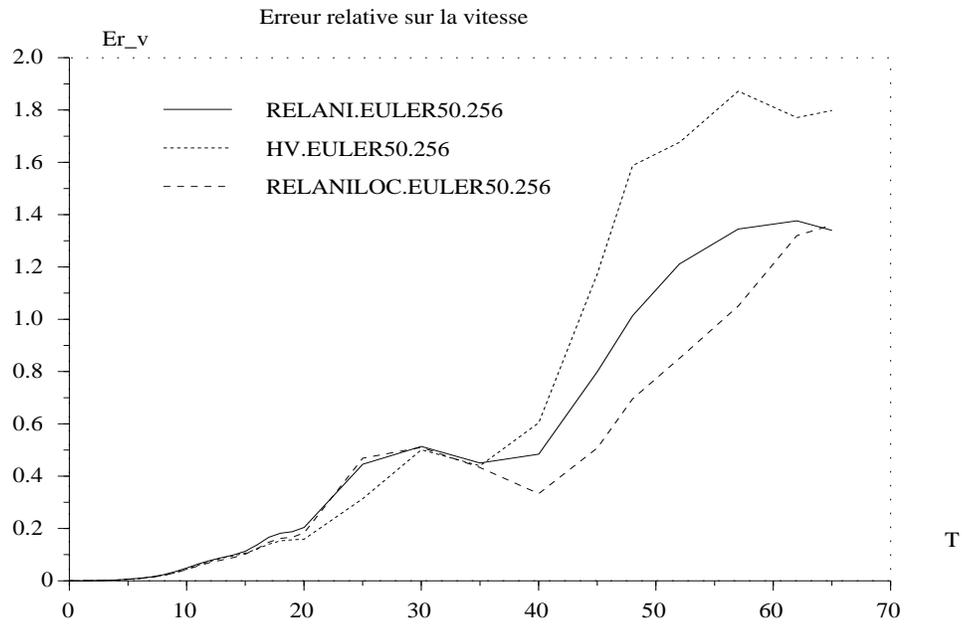}} \par}

\caption{\label{fig: Compare.Euler50.ani.HV.v}Velocity relative error versus
time, for three low resolution computations (256x256), with different
small scale turbulence parameterization: anisotropic relaxation equations
(RELANI.Euler50.256), hyperviscous model (HV.Euler50.256) and anisotropic
relaxation equations with local energy conservation (RELANILOC.Euler50.256).
The initial condition is made of 50 vorticity patches with random
positions. The reference computation is a 1024X1024 resolution hyperviscous
one (HV.EULER50.1024)}
\selectlanguage{english}
\end{figure}

Fig. \ref{fig: Compare.Euler50.ani.HV.v} shows the evolution of the
velocity relative error for three low resolution (256X256) computations:
hyperviscous (HV.Euler50.256), anisotropic relaxation (RELANI.Euler50.256)
and anisotropic relaxation with local energy conservation (RELANILOC.Euler50.256).
We use as a reference an hyperviscous computation with resolution
1024X1024. For smaller times, the three computations give equivalent
results: a quasi-exponential evolution of the error, probably associated
to the intrinsic instability of the initial condition. However, after
a stage of several vortex merging, the anisotropic relaxation equations
give much better results than the hyperviscous ones. This is illustrated
by Fig.s \ref{fig: Compare.Euler50.ani.HV.v} and \ref{Compare_Plot_Euler50_2}.
This last figure show that the position and the structures of the
two final vortices is better reproduced. We propose the following
interpretation: whereas the vortex motion is very sensitive to the
initial condition, the result of the vortex merging depends less on
the detail initial condition. It rather depends on the dynamical invariants
(the global vorticity, the energy). The relaxation equation which
explicitly conserve the energy, and which describes the vorticity
mixing in accordance with the statistical properties of the system
lead to better results than the hyperviscous or the other parameterization
studied here. We thus observe a \emph{statistical saturation} of the
error during this stage. 

These results show that the Navier-Stokes equation is very clearly
the poorer model of two dimensional decaying turbulence. To compute
the error, we have used, as a reference, an hyperviscous type computation.
In order for the results to be non ambiguous, we have shown in \cite{Bouchet_these}
that all the conclusions we give here are also valid when a comparison
is made with Navier-Stokes high resolution computations (HV.Euler50.1024)
(see also figure \ref{Compare_Plot_Euler50_2}).\\

We conclude that in any situations, hyperviscous parameterizations
and anisotropic relaxation equations lead to much better results than
Navier-Stokes equation or the Smagorinsky model. The difference between
hyperviscous and anisotropic relaxation equations computations is
less important. However for small resolution (HV.Euler4.128) or for
very complex flow evolution, the anisotropic relaxation equations
give better results than the hyperviscous parameterization. 

Our validation by comparison with higher resolution computation, does
not allow us to compare different models for very large times. However,
we think that the relaxation equation will then give much better results
than the hyperviscous ones, because they exactly conserve the energy,
because they treat vorticity mixing as far as possible according to
the physical processes responsible for this mixing, and because they
converge towards equilibrium states of the Euler equation. On the
contrary hyperviscous model are not based on any physical ground.

We conclude by stating that similar results have been obtained in
the context of the Quasi-Geostrophic dynamics (not reported, see \cite{Bouchet_these}).
Due to the slowing down of the dynamical mixing of the potential vorticity
for small Rossby deformation radius \( R \) (the interaction is screened
at a typical scale \( R \)), the relative efficiency of the anisotropic
relaxation equations is even more evident in such a case.

\section{Discussion \label{sec_Discussion}}

For geophysical flows, the scale at which the microscopic diffusion
is relevant is so small that any numerical model of oceans or atmosphere
implicitly assumes a small scale parameterization. In such physical
situations, lots of phenomenon occur at the unresolved scales: convection,
interaction with boundary, three dimensional turbulence at smaller
scales, leading to molecular diffusion, etc ... For this reason, the
fundamental equation to start with to describe such flows is not even
clear \cite{Pedlosky}. It is however generally recognized that the
dynamics of geophysical flows is two-dimensional like (inverse energy
cascade) and that the dynamics of the potential vorticity plays a
major role. In order to parameterize the potential vorticity mixing,
we have considered the simplest model having the main properties of
large scale geophysical flows: the barotropic Quasi-geostrophic model.
As the equations are formally equivalent, we have led our study mainly
in the 2D incompressible Euler-equation framework. All the results
obtained are valid for both of these models, and may be straightforwardly
generalized to the multi-layered Quasi-Geostrophic model. The problem
we have addressed is to describe the evolution of the coarse-grained
vorticity, the fundamental equation being the Euler (or QG) equations.
In this study, we have considered a regime of freely decaying turbulence,
with vortex merging and filamentation. As we mainly study this transient
behavior, this study also apply for the two-dimensional Navier-Stokes
equations (or viscous QG) with a molecular viscosity acting on a scale
much smaller than the resolved one.

In section \ref{sec: moyenis=E9e}, we have first defined the coarse-graining.
We have then shown that the parameterization problem is equivalent
to the determination of a turbulent current. We have obtained at leading
order a deterministic approximation of the turbulent current: the
action of the strain tensor on the gradient of the vorticity\cite{Dubos_CRAS,Dubos_Thèse,Bouchet_these}.
We have shown using numerically obtained vorticity fields that this
approximation is a good one in the regime of vortex merging and filamentation.
The leading contribution comes from the part of the turbulent current
which is absent in the case of an ensemble average (like in usual
Reynolds tensor). This shows that the turbulent current is dominated
by systematic correlation between resolved and unresolved scale. This
essentially deterministic process is qualitatively associated to structures
(filaments) passing from resolved to unresolved scales). This deterministic
approximation exactly conserves the energy.

This deterministic approximation acts on the vorticity field as a
positive viscosity in the stretching direction of the strain tensor,
and as a negative viscosity in the expansion direction. Due to this
last contribution, a numerical model using directly this approximation
would lead to numerical instabilities. In order to obtain stable equations,
respecting the actual diffusion in the contraction direction, and
the energy conservation, we look for equations maximizing the mixing
entropy for the vorticity, while imposing these dynamical constraints.
We obtain \emph{anisotropic relaxations equations} which have the
further property to relax towards the statistical equilibrium of the
Euler-equation. 

Some entropy based parameterization of the turbulent current have
been proposed previously in the case of topography dominated flows,
by Holloway (see \cite{Holloway} for a review or \cite{Alvarez_Holloway}
for some application in the context of oceanographic flows). The maximum
entropy production principle has been proposed in \cite{Robert_Sommeria_92},
but an isotropic turbulent current was then imposed. The maximum entropy
production principle has no theoretical or physical ground. The anisotropic
relaxation equations do not for instance describe correctly what happens
in the strain tensor stretching direction. We use here this principle
as a trick to obtain stable equations, with given constraints, and
respecting the statistical tendency of the vorticity mixing. 

In order to show the interest of the anisotropic relaxation equations,
we have compared its efficiency with the Navier-Stokes equations,
or other usual parameterizations such as the Smagorinsky model or
the hyperviscous model. We have compared the results with higher resolution
Navier-Stokes or hyperviscous computations. As they overestimate strongly
the diffusion, the Navier-Stokes or the Smagorinsky model give clearly
the worst results. The anisotropic relaxation equations give better
results than the hyperviscous model, for complex flow evolution, for
smaller resolutions, or for the Quasi-Geostrophic model. 

The results we have obtained clearly show that the anisotropic relaxation
equations allows an increase of the predictability with respect to
other parameterizations. We note however that we have suppose a complete
knowledge of the initial condition. The real problem of predictability
for geophysical flows is much more complex, as lots of error sources
have to be taken into account. We have studied only the effect of
the parameterization of the potential vorticity mixing. \\

Let us conclude on a discussion on the limitations of this work and
on its possible interest in a wider context. We have First we have
studied the regime of vortex merging and strong filamentation. However,
for inviscid two-dimensional freely decaying turbulence, on the latter
stage of the evolution of the flow, the coarse-grained field is close
to a stationary state. In this different regime, no more structures
are transported from larger to smaller scales (scale separation).
In such a case, the main contribution to the turbulent current probably
comes from the stochastic part of the current. The effect of these
fluctuations is a very slow modification of the structure of the quasi-stationary
coarse-grained flow. An ensemble average is probably the best way
to address such a relaxation. We note the application of a quasi-linear
theory in \cite{Chavanis_Quasilinear} which may be of interest for
this regime. 

In the situation we have studied, because the processes are very rapid,
the effect of a very small viscosity (acting on scales much smaller
than the resolution one) is actually negligible. However, in the stage
of a quasi-stationary state relaxation, the effect of a very small
viscosity would probably be very important, as it should determine
the distribution of the unresolved scale vorticity. \foreignlanguage{english}{A
statistical mechanics based approach is proposed in \cite{Grote_Majda}.
There a crude closure,} using the simplest energy-enstrophy statistical
theory for flow with topography, is proposed. In \cite{kaz98} the
effect of a small forcing and dissipation is also considered. \foreignlanguage{english}{}

Another improvement of our work may be done by the description, not
only of the averaged but also of the higher vorticity moments. In
\cite{Robert_Rosier*} an equation hierarchy is derived, to describe
these moment evolution. This hierarchy is then closed using a maximum
entropy production principle. Such ideas have been applied in the
context of a barotropic ocean model \cite{kaz98}.\\

The current computational power allows to make very high resolution
simulations of the two-dimensional Navier Stokes equations. In such
a case, the precise simulation of rapid physical phenomena does not
necessitate any parameterization. However, for more complex physical
situations, for very long time simulations or for more complex geophysical
flow models, the necessity to use good small scales parameterizations
is evident. Let us give some examples. In the modeling of Jupiter's
troposphere, we have shown that the very small value of the Rossby
deformation radius renders the potential vorticity dynamics very slow
\cite{Bouchet_Dumont}. The very long time then needed, for the flow
to organize, does not allow to obtain the characteristic very strong
jets typical of Jupiter's vortices, when using Navier-Stokes equations.
On the contrary, using relaxation equations permits to model these
features very precisely, even using quite low resolution models \cite{Bouchet_Dumont}.
For ocean modeling, let us discuss the example of the Zapiola anticyclone,
in the Argentina basin, in South Atlantic. In the late 90's, Topeix-Poseidon
data has clearly revealed this very strong anticyclonic structure,
whereas up to date South Atlantic modeling where not able to reproduce
it, or were reproducing a structure much weaker than the observed
ones \cite{DeMiranda_Barnier_Zapiola}. In \cite{DeMiranda_Barnier_Zapiola},
for instance, the anticyclone modeling is convincing, using the up
to date model resolution. Moreover, for global ocean models, or/and
when much longer computational times are needed, for instance for
climate modeling, the correct description of such structures would
necessitate unavailable computational power. Moreover, as shown by
Fig. \ref{Compare_Plot_Euler50_2} of this study, for very long time
scales, the use of rough parameterization alters the fluctuations
of vorticity, or potential vorticity. 

The modeling of geophysical flows actually require the best possible
parameterization. The analysis we have done could be straightforwardly
generalized to the case of multi-layered Quasi-geostrophic models.
For z-coordinates models, or isopicnal models in use in geophysical
flow modeling, the generalization of the ideas developed in this article
should be considered ; the problem being much more simple in models
clearly identifying the potential vorticity as a key dynamical variable.
We conclude by saying that the modeling of the potential vorticity
mixing in geophysical fluid dynamics is only one part of the parameterization
problem, but probably one of the more essential one.

\section*{Acknowledgements}

This work has been done during my PHD thesis \cite{Bouchet_these},
supervised by R. Robert and J. Sommeria. I thank M.L. Chabanol, P.H.
Chavanis and B. Dubrulle for useful comments on this work. It has
been partly supported by a \emph{Bourse Lavoisier} of the French \emph{Ministère
des Affaires Etrangères} and by the \foreignlanguage{english}{contract
COFIN00 \emph{Chaos and Localization in Classical and Quantum mechanics.}}
\selectlanguage{english}

\selectlanguage{french}
\newpage

\begin{table}
{\centering \begin{tabular}{|c|c|c|c|}
\hline 
&
\selectlanguage{french}
\( \left\Vert \cdot \right\Vert _{2} \)
\selectlanguage{english}&
\selectlanguage{french}
\( \left\Vert \cdot \right\Vert _{\infty } \)
\selectlanguage{english}&
\selectlanguage{french}
\( \left\Vert \overline{\omega }\right\Vert _{1}/\left\Vert \cdot \right\Vert _{1} \)
\selectlanguage{english}\\
\hline
\hline 
\selectlanguage{french}
\( \nabla .{\bf J}_{\omega } \)
\selectlanguage{english}&
\selectlanguage{french}
1.14
\selectlanguage{english}&
\selectlanguage{french}
1.39
\selectlanguage{english}&
\selectlanguage{french}
7.63
\selectlanguage{english}\\
\hline 
\selectlanguage{french}
\( \nabla .\left( \overline{\overline{\omega }\, \overline{{\bf u}}}-\overline{\omega }\, \overline{{\bf u}}\right)  \)
\selectlanguage{english}&
\selectlanguage{french}
5.34
\selectlanguage{english}&
\selectlanguage{french}
7.16
\selectlanguage{english}&
\selectlanguage{french}
1.64
\selectlanguage{english}\\
\hline 
\selectlanguage{french}
\( \nabla .\left( \overline{\tilde{\omega }\overline{{\bf u}}}\right)  \)
\selectlanguage{english}&
\selectlanguage{french}
5.38
\selectlanguage{english}&
\selectlanguage{french}
7.20
\selectlanguage{english}&
\selectlanguage{french}
1.64
\selectlanguage{english}\\
\hline 
\selectlanguage{french}
\( \nabla .\left( \overline{\overline{\omega }\widetilde{{\bf u}}}\right)  \)
\selectlanguage{english}&
\selectlanguage{french}
0.04
\selectlanguage{english}&
\selectlanguage{french}
0.06
\selectlanguage{english}&
\selectlanguage{french}
229
\selectlanguage{english}\\
\hline 
\selectlanguage{french}
\( \nabla .\left( \overline{\tilde{\omega }\widetilde{{\bf u}}}\right)  \)
\selectlanguage{english}&
\selectlanguage{french}
0.12
\selectlanguage{english}&
\selectlanguage{french}
0.19
\selectlanguage{english}&
\selectlanguage{french}
76.4
\selectlanguage{english}\\
\hline 
\selectlanguage{french}
\( \nabla .{\bf J}_{d} \)
\selectlanguage{english}&
\selectlanguage{french}
1.12
\selectlanguage{english}&
\selectlanguage{french}
1.35
\selectlanguage{english}&
\selectlanguage{french}
7.81
\selectlanguage{english}\\
\hline 
\selectlanguage{french}
\( \nabla .{\bf J}_{s} \)
\selectlanguage{english}&
\selectlanguage{french}
0.08
\selectlanguage{english}&
\selectlanguage{french}
0.14
\selectlanguage{english}&
\selectlanguage{french}
111
\selectlanguage{english}\\
\hline
\end{tabular}\par}

\caption{\label{tab:flux}Comparison of the contribution of the divergence
of the components of the turbulent current, evaluated with several
norms \protect\( \left\Vert f\right\Vert _{2}=\sqrt{1/\left| D\right| \int _{D}d{\bf r}\, f^{2}}\protect \),
\protect\( \left\Vert \cdot \right\Vert _{\infty }={\rm Sup}_{D}\left\{ \left| f\right| \right\} \protect \),
\protect\( \left\Vert f\right\Vert _{1}=1/\left| D\right| \int _{D}d{\bf r}\, \left| f\right| \protect \).
The divergence of the turbulent current \protect\( \nabla .{\bf J}_{\omega }=\nabla .{\bf J}_{d}+\nabla .{\bf J}_{s}\protect \)
is dominated by the contribution of its deterministic part \protect\( \nabla .{\bf J}_{d}\protect \).
The stochastic part \protect\( \nabla .{\bf J}_{s}=\nabla .\left( \overline{\overline{\omega }\widetilde{{\bf u}}}\right) +\nabla .\left( \overline{\tilde{\omega }\widetilde{{\bf u}}}\right) \protect \)
is dominated by \protect\( \nabla .\left( \overline{\tilde{\omega }\widetilde{{\bf u}}}\right) .\protect \)
This last term would be the only contribution in the case of an ensemble
average.}
\selectlanguage{english}
\end{table}
 
\begin{table}
{\centering \begin{tabular}{|c|c|c|c|}
\hline 
&
\selectlanguage{french}
\( \left\Vert \cdot \right\Vert _{2} \)
\selectlanguage{english}&
\selectlanguage{french}
\( \left\Vert \cdot \right\Vert _{\infty } \)
\selectlanguage{english}&
\selectlanguage{french}
\( \left\Vert \overline{\omega }\right\Vert _{1}/\left\Vert \cdot \right\Vert _{1} \)
\selectlanguage{english}\\
\hline
\hline 
\selectlanguage{french}
\( \nabla .{\bf J}_{d} \)
\selectlanguage{english}&
\selectlanguage{french}
1.12
\selectlanguage{english}&
\selectlanguage{french}
1.35
\selectlanguage{english}&
\selectlanguage{french}
7.81
\selectlanguage{english}\\
\hline 
\selectlanguage{french}
\( \nabla .\left( {\bf J}_{d}-l^{2}{\bf \Sigma }_{sym}{\bf \nabla }\overline{\omega }\right)  \)
(ordre 1)
\selectlanguage{english}&
\selectlanguage{french}
0.13
\selectlanguage{english}&
\selectlanguage{french}
0.18
\selectlanguage{english}&
\selectlanguage{french}
69.0
\selectlanguage{english}\\
\hline 
\selectlanguage{french}
\( \nabla .\left( {\bf J}_{d}-l^{2}{\bf \Sigma }{\bf \nabla }\overline{\omega }-{\rm ordre\, \, 2}\right)  \)
\selectlanguage{english}&
\selectlanguage{french}
0.05
\selectlanguage{english}&
\selectlanguage{french}
0.10
\selectlanguage{english}&
\selectlanguage{french}
163
\selectlanguage{english}\\
\hline 
\selectlanguage{french}
\( \nabla .\left( l^{2}{\bf \Sigma }^{<}_{sym}{\bf \nabla }\overline{\omega }\right)  \)
\selectlanguage{english}&
\selectlanguage{french}
1.36
\selectlanguage{english}&
\selectlanguage{french}
1.91
\selectlanguage{english}&
\selectlanguage{french}
6.61
\selectlanguage{english}\\
\hline 
\selectlanguage{french}
\( \nabla .\left( l^{2}{\bf \Sigma }^{>}_{sym}{\bf \nabla }\overline{\omega }\right)  \)
\selectlanguage{english}&
\selectlanguage{french}
0.86
\selectlanguage{english}&
\selectlanguage{french}
1.13
\selectlanguage{english}&
\selectlanguage{french}
10.8
\selectlanguage{english}\\
\hline 
\selectlanguage{french}
\( \nabla .\left( {\bf J}_{d}-l^{2}{\bf \Sigma }^{<}_{sym}{\bf \nabla }\overline{\omega }\right)  \)
\selectlanguage{english}&
\selectlanguage{french}
0.87
\selectlanguage{english}&
\selectlanguage{french}
1.13
\selectlanguage{english}&
\selectlanguage{french}
10.8
\selectlanguage{english}\\
\hline
\end{tabular}\par}

\caption{\label{tab:flux_cisaillement} Norm of the difference between the
divergence of the deterministic part of the turbulent current \protect\( \nabla .{\bf J}_{d}\protect \)
and its first and second order approximations. The leading order approximation
of \protect\( \nabla .{\bf J}_{d}\protect \) is correct up to an
error of order 10\%, for the three norms considered. This error is
of the same order as the one corresponding to the stochastic part
\protect\( \nabla .{\bf J}_{s}\protect \)}
\selectlanguage{english}
\end{table}

\begin{table}
{\centering \begin{tabular}{|c|c|c|c|c|}
\hline 
\selectlanguage{french}
Calcul
\selectlanguage{english}&
\selectlanguage{french}
Résolution
\selectlanguage{english}&
\selectlanguage{french}
\( \nu  \) ou \( l \)
\selectlanguage{english}&
\selectlanguage{french}
\( dt \)
\selectlanguage{english}&
\selectlanguage{french}
\( Re \)
\selectlanguage{english}\\
\hline
\hline 
\selectlanguage{french}
DNS.Euler4.1024
\selectlanguage{english}&
\selectlanguage{french}
1024x1024
\selectlanguage{english}&
\selectlanguage{french}
\( 1.57\, 10^{-5} \)
\selectlanguage{english}&
\selectlanguage{french}
\( 6.14\, 10^{-4} \)
\selectlanguage{english}&
\selectlanguage{french}
\( 400\, 000 \)
\selectlanguage{english}\\
\hline 
\selectlanguage{french}
REL.Euler4.256
\selectlanguage{english}&
\selectlanguage{french}
256x256
\selectlanguage{english}&
\selectlanguage{french}
\( 6.28\, 10^{-4} \)
\selectlanguage{english}&
\selectlanguage{french}
\( 2.45\, 10^{-3} \)
\selectlanguage{english}&
\selectlanguage{french}
\( 10\, 000 \)
\selectlanguage{english}\\
\hline 
\selectlanguage{french}
RELVAR.Euler4.256
\selectlanguage{english}&
\selectlanguage{french}
256x256
\selectlanguage{english}&
\selectlanguage{french}
\( 1.26\, 10^{-4} \)
\selectlanguage{english}&
\selectlanguage{french}
\( 2.45\, 10^{-3} \)
\selectlanguage{english}&
\selectlanguage{french}
-
\selectlanguage{english}\\
\hline 
\selectlanguage{french}
RELANI.Euler4.256
\selectlanguage{english}&
\selectlanguage{french}
256x256
\selectlanguage{english}&
\selectlanguage{french}
\( 1.33\, 10^{-2} \)
\selectlanguage{english}&
\selectlanguage{french}
\( 2.45\, 10^{-3} \)
\selectlanguage{english}&
\selectlanguage{french}
-
\selectlanguage{english}\\
\hline 
\selectlanguage{french}
RELANILOC.Euler4.256
\selectlanguage{english}&
\selectlanguage{french}
256x256
\selectlanguage{english}&
\selectlanguage{french}
\( 1.33\, 10^{-2} \)
\selectlanguage{english}&
\selectlanguage{french}
\( 2.45\, 10^{-3} \)
\selectlanguage{english}&
\selectlanguage{french}
-
\selectlanguage{english}\\
\hline 
\selectlanguage{french}
DNS.Euler4.256
\selectlanguage{english}&
\selectlanguage{french}
256x256
\selectlanguage{english}&
\selectlanguage{french}
\( 6.28\, 10^{-4} \)
\selectlanguage{english}&
\selectlanguage{french}
\( 2.45\, 10^{-3} \)
\selectlanguage{english}&
\selectlanguage{french}
\( 10\, 000 \)
\selectlanguage{english}\\
\hline 
\selectlanguage{french}
SMA.Euler4.256
\selectlanguage{english}&
\selectlanguage{french}
256x256
\selectlanguage{english}&
\selectlanguage{french}
\( 1.33\, 10^{-2} \)
\selectlanguage{english}&
\selectlanguage{french}
\( 2.45\, 10^{-3} \)
\selectlanguage{english}&
\selectlanguage{french}
\( - \)
\selectlanguage{english}\\
\hline 
\selectlanguage{french}
HV.Euler4.256
\selectlanguage{english}&
\selectlanguage{french}
256x256
\selectlanguage{english}&
\selectlanguage{french}
\( 1.92\, 10^{-8} \)
\selectlanguage{english}&
\selectlanguage{french}
\( 2.45\, 10^{-3} \)
\selectlanguage{english}&
\selectlanguage{french}
\( - \)
\selectlanguage{english}\\
\hline 
\selectlanguage{french}
RELANI.Euler4.128
\selectlanguage{english}&
\selectlanguage{french}
128x128
\selectlanguage{english}&
\selectlanguage{french}
\( 2.45\, 10^{-2} \)
\selectlanguage{english}&
\selectlanguage{french}
\( 4.91\, 10^{-3} \)
\selectlanguage{english}&
\selectlanguage{french}
-
\selectlanguage{english}\\
\hline 
\selectlanguage{french}
HV.Euler4.128
\selectlanguage{english}&
\selectlanguage{french}
128x128
\selectlanguage{english}&
\selectlanguage{french}
\( 1.53\, 10^{-7} \)
\selectlanguage{english}&
\selectlanguage{french}
\( 4.91\, 10^{-3} \)
\selectlanguage{english}&
\selectlanguage{french}
\( - \)
\selectlanguage{english}\\
\hline
\end{tabular}\par}

\caption{\label{table:parametres_Euler}Numerical parameters for the computations
of the coalescence of four vortices}
\selectlanguage{english}
\end{table}
 
\begin{table}
{\centering \begin{tabular}{|c|c|c|c|c|}
\hline 
\selectlanguage{french}
Calcul
\selectlanguage{english}&
\selectlanguage{french}
Résolution
\selectlanguage{english}&
\selectlanguage{french}
\( \nu  \) ou \( l \)
\selectlanguage{english}&
\selectlanguage{french}
\( dt \)
\selectlanguage{english}&
\selectlanguage{french}
\( Re \)
\selectlanguage{english}\\
\hline
\hline 
\selectlanguage{french}
DNS.Euler50.1024
\selectlanguage{english}&
\selectlanguage{french}
1024x1024
\selectlanguage{english}&
\selectlanguage{french}
\( 3.14\, 10^{-5} \)
\selectlanguage{english}&
\selectlanguage{french}
\( 6.14\, 10^{-4} \)
\selectlanguage{english}&
\selectlanguage{french}
\( 200\, 000 \)
\selectlanguage{english}\\
\hline 
\selectlanguage{french}
DNS.Euler50.1024
\selectlanguage{english}&
\selectlanguage{french}
1024x1024
\selectlanguage{english}&
\selectlanguage{french}
\( 5.99\, 10^{-11} \)
\selectlanguage{english}&
\selectlanguage{french}
\( 6.14\, 10^{-4} \)
\selectlanguage{english}&
\selectlanguage{french}
-
\selectlanguage{english}\\
\hline 
\selectlanguage{french}
RELANI.Euler50.256
\selectlanguage{english}&
\selectlanguage{french}
256x256
\selectlanguage{english}&
\selectlanguage{french}
\( 1.33\, 10^{-2} \)
\selectlanguage{english}&
\selectlanguage{french}
\( 2.45\, 10^{-3} \)
\selectlanguage{english}&
\selectlanguage{french}
-
\selectlanguage{english}\\
\hline 
\selectlanguage{french}
RELANILOC.Euler50.256
\selectlanguage{english}&
\selectlanguage{french}
256x256
\selectlanguage{english}&
\selectlanguage{french}
\( 1.33\, 10^{-2} \)
\selectlanguage{english}&
\selectlanguage{french}
\( 2.45\, 10^{-3} \)
\selectlanguage{english}&
\selectlanguage{french}
-
\selectlanguage{english}\\
\hline 
\selectlanguage{french}
HV.Euler50.256
\selectlanguage{english}&
\selectlanguage{french}
256x256
\selectlanguage{english}&
\selectlanguage{french}
\( 1.92\, 10^{-8} \)
\selectlanguage{english}&
\selectlanguage{french}
\( 2.45\, 10^{-3} \)
\selectlanguage{english}&
\selectlanguage{french}
\( - \)
\selectlanguage{english}\\
\hline
\end{tabular}\par}

\caption{\label{table:parametres_Euler50}Numerical parameters for the various
parameterization of the coalescence of 50 vorticity patches. }
\selectlanguage{english}
\end{table}

\selectlanguage{english}

\end{document}